\documentclass{jfm}

            \newif\ifdraft \newif\ifcolorfigs \newif\ifJFM
            \drafttrue \colorfigstrue \JFMtrue

\ifJFM
  \usepackage{upmath}
\fi

\usepackage{ifpdf}
\ifpdf
  \usepackage[pdftex]{color,graphicx}
  \usepackage{epstopdf}
\else
  \usepackage[dvips]{color,graphicx}
\fi

\usepackage{float}
\usepackage{natbib}
\usepackage{amsmath,amsfonts,amssymb,amsbsy,amscd,amsgen}
\usepackage{url}
\usepackage{subfigure}
\usepackage{ifthen}
\graphicspath{{../figs/}}  

\usepackage[active]{srcltx}

\ifcolorfigs
  \usepackage[pdftex,colorlinks]{hyperref} 
\else
\fi

\ifdraft    
   
   \newcommand{\PC}[1]{$\footnotemark\footnotetext{PC: #1}$}
   
   \newcommand{\MA}[1]{$\footnotemark\footnotetext{MA: #1}$}
   
   \newcommand{\APW}[1]{$\footnotemark\footnotetext{APW: #1}$}

   \newcommand{\file}[1]{$\footnotemark\footnotetext{{\bf file} #1}$}
   \newcommand{\mycomment}[2]{\noindent \textbf{\underline{#1}}: \emph{#2}}
\else   

   \newcommand{\PC}[1]{}
   \newcommand{\JFG}[1]{}
   \newcommand{\MA}[1]{}
   \newcommand{\APW}[1]{}

   \newcommand{\file}[1]{}
   \newcommand{\mycomment}[2]{}

\fi 

\ifcolorfigs            

  \newcommand{\HREF}[2]{{\href{#1}{#2}}}

\else

  \newcommand{\HREF}[2]{{#2}}

\fi

\newcommand{\refref} [1] {\cite{#1}}

\newcommand{\refeq}  [1] {(\ref{#1})}

\newcommand{\reffig} [1] {figure~\ref{#1}}

\newcommand{\refFig} [1] {Figure~\ref{#1}}

\newcommand{\reftab} [1] {table~\ref{#1}}

\newcommand{\refsect}[1] {\S\,\ref{#1}}
\newcommand{\refsects}[2] {\S\,\ref{#1} and \S\,\ref{#2}}

\newcommand{\refappe}[1] {appendix~\ref{#1}}

\newcommand{\beq}{\begin{equation}}
\newcommand{\continue}{\nonumber \\ }

\newcommand{\eeq}{\end{equation}}
\newcommand{\ee}[1] {\label{#1} \end{equation}}

\newcommand{\bea}{\begin{eqnarray}}
\newcommand{\eea}{\end{eqnarray}}
\newcommand{\barr}{\begin{array}}
\newcommand{\earr}{\end{array}}

\newcommand{\ie}{{i.e.}}        
\newcommand{\eg}{{e.g.}}        

\ifJFM                          

\renewcommand\eg{e.g.\ }
\else
\fi

\newcommand{\gSpace}{\ensuremath{{\bf \phi}}}   
\newcommand{\velRel}{\ensuremath{c}}    
\newcommand{\phaseVel}{phase velocity}      

\newcommand\PoincSec{Poincar\'e section}

\newcommand{\braket}[2]
		   {\langle{#1}\vphantom{#2}|\vphantom{#1}{#2}\rangle}

\newcommand{\sspRSing}{\ensuremath{\sspRed^*}} 	

\newcommand{\po}{periodic orbit}

\newcommand{\rpo}{rela\-ti\-ve periodic orbit}

\newcommand{\reducedsp}{reduced state space}

\newcommand{\slice}{slice}
\newcommand{\Slice}{Slice}
\newcommand{\mslices}{method of slices}
\newcommand{\Mslices}{Method of slices}

\newcommand{\chartBord}{chart border}

\newcommand{\template}{template} 
\newcommand{\sliceBord}{slice border}

\newcommand{\zeit}{\ensuremath{t}}  
\newcommand{\normVec}{\ensuremath{\mathbf{n}}}    
\newcommand{\sliceTan}[1]{\ensuremath{\mathbf{t}{}'_{#1}}}    
\newcommand{\groupTan}{\ensuremath{\mathbf{t}}}    
\newcommand{\Lg}{\ensuremath{\mathbf{T}}}   
\newcommand{\LieEl}{\ensuremath{g}}  
\newcommand{\id}{{\ \hbox{{\rm 1}\kern-.6em\hbox{\rm 1}}}}

\newcommand{\On}[1]{\ensuremath{\textrm{O}(#1)}}
\newcommand{\SOn}[1]{\ensuremath{\textrm{SO}(#1)}}         
\newcommand{\Dn}[1]{\ensuremath{\textrm{D}_{#1}}}              
\newcommand{\Zn}[1]{\ensuremath{\textrm{C}_{#1}}}              

\newcommand{\pSRed}{\ensuremath{\hat{\cal M}}} 
\newcommand{\sspRed}{\ensuremath{\hat{\ssp}}}    
\newcommand{\velRed}{\ensuremath{\hat{\vel}}}    
\newcommand{\slicep}{\ensuremath{\ssp'}}   
\newcommand{\Group}{\ensuremath{G}}         

\newcommand{\NS}{Navier--Stokes}
\newcommand{\NSe}{Navier--Stokes equations}

\newcommand{\KS}{Kuramoto--Sivashinsky}
\newcommand{\Reynolds}{\textit{Re}}  
\newcommand{\pCf}{plane Couette flow}

\newcommand{\eqv}{equilib\-rium}

\newcommand{\eqva}{equilib\-ria}

\newcommand{\reqv}{travelling wave}

\newcommand{\reqva}{travelling waves}
\newcommand{\Reqva}{Travelling waves}

\newcommand{\cohStr}{coherent structure}
\newcommand{\recurrStr}{recurrent coherent structure}

\newcommand{\stateDsp}{state-space}
\newcommand{\StateDsp}{State-space}

\newcommand{\statesp}{state space}

\newcommand{\be}{{\bf e}}
\newcommand{\bu}{{\bf u}}
\newcommand{\bx}{{\bf x}}
\newcommand{\Gpipe}{\ensuremath{\Gamma}} 

\newcommand{\REQV}[2]{\ensuremath{\mathrm{TW}_{#1#2}}} 

\newcommand{\RPO}[1]{\ensuremath{\mathrm{RPO}_{#1}}}

\newcommand{\bCell}{\ensuremath{\Omega}}
\newcommand{\Norm}[1]{\|{#1}\|}

\newcommand{\dmn}{-dimensional} 

\newcommand{\timeAver} [1]{\overline{#1}}

\newcommand{\shift}{\ensuremath{\ell}}
\newcommand\period[1]{{T_{#1}}}         
\newcommand{\pS}{{\cal M}}          
\newcommand{\ssp}{a}            
\newcommand{\vel}{\ensuremath{v}}   

\newcommand{\eigExp}[1][]{
\ifthenelse{\equal{#1}{}}{\ensuremath{\lambda}}{\ensuremath{\lambda^{(#1)}}}
                        }
\newcommand{\eigRe}[1][]{
\ifthenelse{\equal{#1}{}}{\ensuremath{\mu}}{\ensuremath{\mu^{(#1)}}}
                        }
\newcommand{\eigIm}[1][]{
  \ifthenelse{\equal{#1}{}}{\ensuremath{\omega}}{\ensuremath{\omega^{(#1)}}}
            }

\newcommand{\bnabla}{\mbox{\boldmath $\nabla$}}
\renewcommand{\vec}[1]{\mbox{\boldmath $#1$}}

\title[Revealing the state space of turbulent pipe flow]
{Revealing the state space of turbulent pipe flow by symmetry reduction}
\author[A.\ P.\ Willis, P.\ Cvitanovi{\'c} and M.\ Avila]
{
A.\ns P.\ns W\ls I\ls L\ls L\ls I\ls S$^1$,
\ns
P.\ns C\ls V\ls I\ls T\ls A\ls N\ls O\ls V\ls I\ls \'C$^2$
\ns
\break
\and
M.\ns A\ls V\ls I\ls L\ls A$^{3,4}$
}
\affiliation{
$^1$School of Mathematics and Statistics,
University of Sheffield, S3\,7RH, U.K.
\\[\affilskip]
$^2$School of Physics,
 Georgia Institute of Technology,
 Atlanta, GA  30332, USA
\\[\affilskip]
$^3$Max Planck Institute for Dynamics and Self-Organization
  (MPIDS),\\ 37077 G\"ottingen, Germany
\\[\affilskip]
$^4$Institute of Fluid Mechanics, Fridriech-Alexander-Universit\"at Erlangen-N\"urnberg,\\
Cauerstrasse 4, 91058 Erlangen, Germany
}

\pubyear{2012}
\begin{document}
\maketitle

Symmetry reduction by the
method of slices is applied to pipe flow in order to quotient the
stream-wise translation and azimuthal rotation symmetries of turbulent
flow states.
Within the
symmetry-reduced state space, all travelling wave solutions reduce to
equilibria, and all relative periodic orbits reduce to periodic orbits.
Projections of these solutions and their unstable manifolds from their
$\infty$-dimensional symmetry-reduced state space onto suitably chosen
2-~or 3-dimensional subspaces reveal their interrelations and the role
they play in organising turbulence in wall-bounded shear flows.
Visualisations of the flow within the slice and its linearisation at
equilibria enable us to trace out the unstable manifolds, determine close
recurrences, identify connections between different travelling wave
solutions, and find, for the first time for pipe flows, relative periodic
orbits that are embedded within the chaotic attractor, which capture
turbulent dynamics at transitional Reynolds numbers.

\section{Introduction}
\label{s:intro}

The understanding of chaotic dynamics in high-dimensional systems that
has emerged in the last decade offers a promising dynamical framework
to study turbulence. Here turbulence is viewed as a walk through a
forest of exact solutions in the $\infty$-dimensional \stateDsp\ of
the governing equations. In pipe flow, the discovery of unstable
\reqva\ \citep{FE03,WK04}, together with glimpses of them in
experiments \citep{science04}, has spurred interest in obtaining a
description of turbulent flow in terms of the dynamics of a handful of
key exact solutions. However, evidence of the relevance of the
dynamical system approach to turbulence has so far been mostly
provided by studies of \pCf\ \citep{GHCW07,HGC08,GHCV08}, with the
discovery of periodic \citep{KawKida05,CviGib10} and relative periodic
orbits \citep{Visw07b} embedded in turbulence playing the key role.
In this approach, the dynamics of turbulent flows at moderate Reynolds
number (\Reynolds) is visualised using \eqv\ solutions of the \NSe\ to
define dynamically invariant, intrinsic, and
representation-independent coordinate frames \citep{GHCW07}.
The resulting visualisations show the role exact solutions
play in shaping turbulence: the observed {\cohStr s} are the physical
images of the flow's least unstable invariant solutions, with
turbulent dynamics arising from a sequence of transitions between
these states. Here the intrinsic low-dimensionality of turbulence
stems from the low number of unstable eigendirections for each state.
In this picture \po s are of particular importance, as they provide
the skeleton underpinning the chaotic dynamics \citep{DasBuch}. In
shear flows evidence is emerging that the geometry of the
\statesp\ near the onset of turbulence is governed by a chaotic
saddle, a set of unstable solutions and their heteroclinic connections
\citep{MuKe05}. The long-term goals of this research program are to
develop this vision into a quantitative, predictive description of
moderate-{\Reynolds} turbulence, and to use this description to
control flows and explain their statistics.

In contrast to \pCf, pipe flow has a non-zero mean axial velocity and
cannot sustain \eqva\ and \po s with both
broken translational symmetry and
zero {\phaseVel}.
Hence in pipes, unstable invariant solutions are generically
stream-wise travelling solutions. The dynamical importance of
invariant solutions is specified by periodic orbit theory, in which
the contribution of each solution to any dynamical average over the
chaotic component of the flow is quantified by a deterministic weight
\citep{DasBuch}. In the presence of continuous symmetries periodic
orbit theory extends to weighted sums over \emph{relative} periodic
orbits \citep{Cvi07}. While a large number of unstable \reqva\ have been
identified in pipe flow \citep{FE03,WK04,Pringle07,Pringle09}, their
neighbourhoods are visited for only $10$-$20\%$ of the time
\citep{KeTu06,SchEckVoll07,WillKer08}, and so it is expected that \rpo
s capture most of the natural measure of the turbulent flow. Although
a few unstable `tiny' \rpo s have already been found
\citep{duguet08,mellibovsky11}, these stem
from bifurcations of nearby \reqva\ and exhibit only minute deviations
about them. More recently, \citet{mellibovsky12} have identified a new
\rpo\ appearing at a global Shilnikov-type bifurcation. All
these solutions, however, lie far from turbulent dynamics and hence
do not provide information about the structure of the chaotic saddle
underlying turbulent flow.

\begin{figure}
\centering
(a)\includegraphics[width=0.45\textwidth,clip=true]{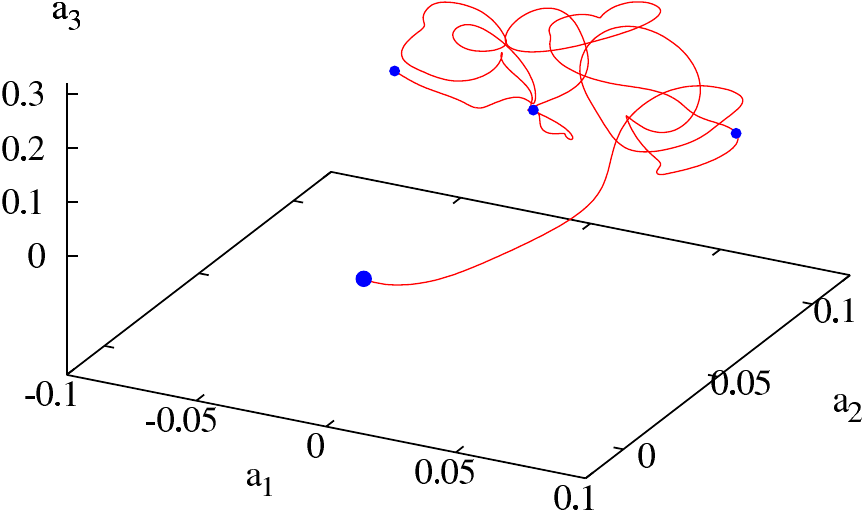}
~~(b)\includegraphics[width=0.45\textwidth,clip=true]{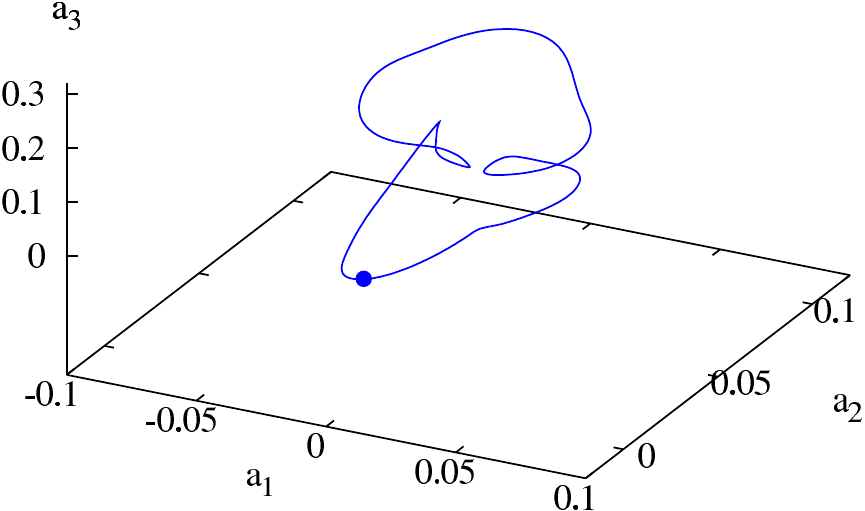}
  \caption{\label{f:MeanVelocityFrame}
Symmetry reduction replaces each full \statesp\ trajectory $\ssp(\zeit)$
by a simpler \reducedsp\ trajectory $\sspRed(\zeit)$, with continuous
group induced drifts quotiented out. Here this is illustrated by the
\rpo\ $\RPO{36.72}$ (see \reffig{fig:M1FULL}\,{\it b})
(a) 
traced in the full {\statesp} for three $\period{}=36.72$ periods, in the
frame moving with the constant axial {\phaseVel}
$\timeAver{\velRel}=1.274$, the average {\phaseVel} of structures estimated
from a long simulation; dots are spaced $\period{}$ apart in time;
(b) 
restricted to the symmetry-\reducedsp. Both are projected onto the
$3$\dmn\ frame \refeq{FrenetFrame1}. In the full \statesp\ a \rpo\ traces
out quasi-periodically a highly contorted 2-torus; in the \reducedsp\ it
closes a \po\ in one period $\period{}$.
            }
\end{figure}

One of the main difficulties in identifying \rpo s embedded in
turbulence is that each of them travels downstream with its own mean
{\phaseVel}. Therefore there is no single co-moving frame that can
simultaneously reduce \emph{all} \rpo s to \po s and all travelling
waves to \eqva. This problem is here addressed by the {\mslices}
\citep{rowley_reconstruction_2000,BeTh04,SiCvi10,FrCv11}, in which the
group orbit of any full-flow structure is represented by a single
point, the group orbit's intersection with a fixed co-dimension one
hypersurface or \emph{`\slice'}. Although this is analogous to the way
a \PoincSec\ reduces a continuous time orbit to a sequence of points,
it should be stressed that a \slice\ is \emph{not} a \PoincSec. A
\slice\ fixes only the group parameters: a continuous time full space
orbit remains a continuous time orbit in the symmetry-\reducedsp\ (see
\reffig{f:MeanVelocityFrame}).

Our goals are two-fold. First, we explain what symmetry reduction is
and how it can aid in revealing the geometry of the \statesp\ of pipe
flow. Second, we demonstrate that this new tool enables us to commence
a systematic exploration of the hierarchy of dynamically important
invariant solutions of pipe flow. Symmetry reduction is here combined
with $3D$ spatial visualisation of instantaneous velocity fields to
elucidate the physical processes underlying the formation of unstable
coherent structures. Running concurrently, the $\infty$-dimensional
\stateDsp\ representation \citep{GHCW07}, enables us to track the
unstable manifolds of invariant solutions, the heteroclinic
connections between them \citep{GHCV08}, and {provides us with} new
insights into the nonlinear \statesp\ geometry and dynamics of
moderate \Reynolds\ wall-bounded flows. Starting in neighbourhoods of
the known \reqva\ \citep{Pringle09} as initial conditions and then
searching for close {recurrences} \citep{pchaot,CviGib10} in the
\reducedsp\ yields educated guesses for locations of \rpo s. Applying
Newton--Krylov methods to these initial guesses leads to the discovery
reported here, the first examples of \rpo s embedded into pipe
turbulence (see \reffig{fig:M1FULL}\,{\it b}).

The paper is organised as follows. We review pipe flows, their
visualisation, and their symmetries in \refsect{s:review}. The
{\mslices} is described in \refsect{s:slice}, and the computation of
invariant solutions and their stability eigenvalues and eigenvectors
in \refsects{s:algorithm}{s:eqbSols}. The main advances reported in
this paper are the symmetry-\reducedsp\ visualisation of
moderate-\Reynolds\ turbulent pipe flow, revealing the
unstable manifolds of \reqva,
and the determination of new
\rpo s\
(\refsect{s:rpos}). Outstanding
challenges are discussed in \refsect{s:concl}. Appendix
\ref{appe:DiscSymmPipe} contains a classification of invariant solutions
according to their symmetries.

\section{Pipe flows}
\label{s:review}

The flow to be considered is that of an incompressible viscous fluid
confined within a pipe of circular cross-section, driven by a constant
mass-flux in the axial direction.  The Reynolds number is defined as
$\Reynolds = U D / \nu$, where $U$ is the mean velocity of the flow,
$D$ is the pipe diameter, and $\nu$ is the kinematic viscosity. We
scale lengths by $D$ and velocities by $U$ in the \NSe\ for $\vec{u}$,
the deviation from the laminar Hagen--Poiseuille flow
\eqv\ $\vec{U}(r)= 2\,(1-2\,r^2)\,\hat{\vec{z}}$,
\beq
\frac{\partial\vec{u}}{\partial t} + \vec{U}\cdot\bnabla\vec{u} +
\vec{u}\cdot\bnabla \vec{U} + \vec{u}\cdot\bnabla\vec{u} = - {\bnabla} p
+ 32\frac{\beta}{\Reynolds}\hat{{\bf z}} +
\frac{1}{\Reynolds}{\bnabla}^2 \vec{u}\,, \qquad \nabla \cdot \vec{u} =
0 \,.
\ee{NavStokesDev}
Hereafter all times will be expressed in dimensionless units $D/U$. Note
that the dimensionless variable $\beta=\beta(\zeit)$ is the fractional
pressure gradient needed to maintain a constant mass-flux, additional
to that required to drive the laminar flow.
A Reynolds number
$\Reynolds_p$, based on the applied pressure gradient, is given by
$\Reynolds_p =\Reynolds \, (1 + \beta)$, whereas the friction Reynolds
number is $\Reynolds_{\tau}=\sqrt{2\Reynolds_p}$. The \NSe\ are
formulated in cylindrical-\-polar coordinates, where $(r, \theta, z)$
are the radius, azimuthal angle and the stream-wise (axial) positions,
respectively. The full fluid velocity field $\vec{U}(r)+\vec{u}$
is represented by $[u,v,w,p](r,\theta,z)$, with $u$, $v$ and
$w$ respectively the radial, azimuthal and stream-wise velocity
components, and $p$ the pressure.

In numerical simulations no-slip boundary conditions are imposed at
the walls and the infinite pipe is represented by periodic boundary
conditions in the stream-wise $z$ direction. Hence the deviation
velocity field $\vec{u}$ and the deviation pressure in the
\NSe\ \refeq{NavStokesDev} are expanded in Fourier modes in the axial
and azimuthal directions,
\beq
\vec{u}(r_n,\theta,z) = \sum_{|k|<K} \sum_{|m'|<M} \vec{u}_{nkm'}
\, \mathrm{e}^{\mathrm{i}(2\alpha kz+m m'\theta)}
\,,
\ee{pipeDiscr}
whereas the finite-difference method is used in the radial direction.
The computational cell is
\beq
\bCell =  [1/2,2\upi/m,\upi/\alpha]
  \equiv  \{(r,\theta,z) \in [0,1/2]\times[0,2\upi/m]\times[0,\upi/\alpha]\}
\,,
\ee{cellPipe}
where $L=\pi/\alpha$ is the
length of the pipe. While $m=1$ corresponds to the naturally periodic
azimuthal boundary condition, \eg  $m=2$ requires that the velocity
field repeats itself twice in $\theta$.  This study is conducted at
\bea
\Reynolds &=& 2400 \,,\; m=2 \,,\; \alpha=1.25
    \continue
\bCell~ &=& [1/2,\upi,\upi/1.25] \,\, \approx \, [90, 283, 452]\;\;\;
\text{wall units}
\,,
    \label{shortPipe}
\eea
corresponding to a short
$L\simeq 2.5D$-periodic pipe in the stream-wise
direction. \cite{mellibovsky11,mellibovsky12} have also focused on
$m=2$ and studied cells with $\alpha\in[1.1,1.85]$. Furthermore, in
this paper we restrict the dynamics to the `shift-and-reflect'
\refeq{ShiftReflOnl} invariant subspace: all invariant solutions and
turbulence simulations presented here are restricted to this subspace.

In this computational cell at $\Reynolds = 2400$ the additional
pressure fraction required to support turbulence while keeping
constant mass-flux is $\timeAver{\beta}=0.70$, yielding friction
Reynolds number $Re_\tau=90.3$. Here one pipe radius $D/2$ corresponds
to about $90$ wall units. At this Reynolds number and geometry
turbulence is found to be transient, with characteristic lifetimes of
order $\zeit \approx 10^3\,D/U$ before the flow finally
relaminarises. It is worth noting that in long pipes without symmetry
restrictions such a characteristic lifetime is found at $\Reynolds=1880$
\citep{hof2008,avila2010}, where the flow takes the form of
stream-wise localised puffs.

The domain size \refeq{shortPipe} was chosen as a compromise between
the computational preference for small domains {\em vs.} the need for
the pipe to be sufficiently long to accommodate turbulent dynamics. In
addition, restricting the largest wavelength is very useful in
identifying key coherent structures characterising turbulent dynamics
\citep{HaKiWa95}. Although the pipes studied in this paper are short,
the three-dimensional states explored here by \eqva\ and their
unstable manifolds are strikingly similar to typical states in longer
pipes.

\subsection{{\StateDsp} visualization of fluid flows}
\label{s:visualStatSp}

As long as one is focusing on a single solution of the \NSe, there are
many excellent, physically insightful $3D$ visualisations of the flow:
velocity fields on flow sections, isovorticity surfaces, videos of the
flow, and so on. But today dozens of exact \eqv\ and \reqv\ solutions
are known for a given turbulent flow, and the number is steadily
growing. Furthermore, we are now commencing an exploration of states
of turbulent fluids in terms of unstable \po\ solutions, whose number
grows exponentially as a function of increasing period. How are we to
visualise \emph{the totality} of these solutions in one go?

The answer was given by \cite{hopf48}. He envisioned the function
space of {\NS} velocity fields as an infinite-dimensional
\statesp\ $\pS$ in which each instantaneous state of $3D$ fluid
velocity field $\vec{u}(\bx)$ is represented as a unique point
$\ssp$. In our particular application we can represent $\ssp =
(\vec{u}_{nkm})$ as a vector whose elements are the primitive
discretization variables \refeq{pipeDiscr}. The $3D$ velocity field
given by $\vec{u}_{knm}(\zeit)$, obtained from integration of the
\NSe\ in time, can hence be seen as trajectory $\ssp(\zeit)$ in
$\approx 100,000$ dimensional space spanned by the free variables of
our numerical discretisation, with the \NS\ equations
\refeq{NavStokesDev} rewritten as
\beq
   \dot{\ssp} = \vel(\ssp) ,
   \qquad
   \ssp(\zeit) = \ssp(0)
            + \int_0^\zeit \! \mathrm{d}\zeit' \, \vel(\ssp(\zeit'))
\,.
\ee{symbolicNS}

Here the current state of the fluid $ \ssp(\zeit)$ is the time-$\zeit$
forward map of the initial fluid state $\ssp(0)$. In order to quantify
whether two fluid states are close to or far from each other, one
needs a notion of distance between two points in \statesp, measured
here as
\beq
  \Norm{\ssp-\ssp'}^2  = \braket{\ssp-\ssp'}{\ssp-\ssp'} =
\frac{1}{V}
\int_\bCell \! d \bx \;
(\vec{u}-\vec{u}') \cdot (\vec{u}-\vec{u}')
\,.
\ee{innerproduct}
There is no compelling reason to use this {`energy norm'}, other than
that velocity fields is what is generated in a numerical
computation. What norm one actually uses depends very much on the
application. For example, in the study of `optimal perturbations' that
move a laminar solution to a turbulent one, both energy
\citep{TeHaHe10} and dissipation \citep{LoCaCoPeGo11} norms have been
used.  In our quest for \reqva\ and \rpo s (see \refsect{s:rpos}) we
find it advantageous to use a `compensatory' norm \refeq{compensNorm}
that enhances the weight of cross-stream velocities.

Visualisations of the \stateDsp\ trajectory \refeq{symbolicNS} are by
necessity projections onto two or three dimensions. Flow states can be
characterised by the instantaneous kinetic energy of their velocity
field, $E = \frac{1}{2} \Norm{\vec{U+u}}^2$, and energy dissipation
rate $D = {\Reynolds}^{-1}\Norm{\bnabla \times (\vec{U}+\vec{u})}^2$.
The dissipation rate is balanced by the energy fed into the flow as
\beq
\dot{E}=I-D
\,,
\ee{Power=I-D}
where
\(
I = \frac{1}{V}
\oint dS \, [{\bf n}\cdot(\vec{u+U})]\, p
\)
is the external power required to maintain constant mass-flux. A
physically appealing choice is to monitor the flow in terms of these
symmetry-invariant, physical observables
$(E(\zeit)/E_\mathrm{lam},D(\zeit)/D_\mathrm{lam},I(\zeit)/I_\mathrm{lam})$,
as in \reffig{fig:M1Orb}. Note that $I(\zeit)/I_\mathrm{lam}=1+\beta(t)$.
For \reqva\ the kinetic energy is constant, so that $D=I$.  Such
solutions sit on the diagonal in \reffig{fig:M1Orb}\,({\it a}), whereas for
\rpo s the kinetic energy is time-periodic, with
$\timeAver{D}=\timeAver{I}$ only for long-time averages.
Whilst this is a good check on correctness of numerical data, such projections
bunch all invariant solutions and turbulent flow along the
energy-balance lines, even though the solutions themselves can be (and
often are) very distant from each other. In fact, if two fluid states
are clearly separated in such plot, they are also separated in the
high-dimensional \statesp. However, the converse is not true; states
of very different topology might have comparable energies, and such
plots may obscure some of the most relevant features of the
flow. Furthermore, relations such as \refeq{Power=I-D} depend on
detailed type and geometry of a given problem
\citep{ksgreene88,SCD07}, and further physical observables beyond
$(E(\zeit),D(\zeit),I(\zeit))$ are difficult to construct.

Recently, \cite{GHCW07} have shown that with the {\statesp}  considered
as a high-dimensional vector space, the dynamics
can be elucidated more profitably by computationally
straight\-forward sets of \emph{physical} coordinates. First, one
identifies several prominent flow states $\vec{u}_A$, $\vec{u}_B$,
$\dots$, such as {\eqva} and their linearised stability eigenvectors,
in whose neighbourhoods the turbulent flow spends most of the
time. From them an orthonormal basis set $\{\be_1, \be_2, \cdots,
\be_n\}$ is constructed by Gram-Schmidt and/or (anti)-symmetrizations. The
evolving fluid state $\bu(\zeit)$ is then projected onto this basis
using the inner product \refeq{innerproduct},
\beq
\ssp(\zeit) =(\ssp_1, \ssp_2, \cdots, \ssp_n, \cdots)(\zeit)
    \,,\qquad
\ssp_n(\zeit) = \braket{\vec{u}(\zeit)}{\be_n}
\,.
\ee{intrSspTraj}
Finally, low-dimensional projections of the flow can be viewed in any
of the $2D$ planes $(\ssp_m, \ssp_n)$ or in $3D$ perspective views
$(\ssp_{\ell},\ssp_m, \ssp_n)$. An example is the
\reffig{f:MeanVelocityFrame} projection on the $3$\dmn\ frame
$\{{\be}_1,{\be}_2,{\be}_3\}$ defined in \refeq{FrenetFrame1}.

It is worth emphasising that this method offers a low-dimensional
{\em visualisation} without dimension reduction or low-dimensional
{\em modelling}; the dynamics are computed with fully-resolved direct
numerical simulations. Although the use of particular \reqva\ to
define low-dimensional projections (see \refsect{s:eqbSols}) may
appear arbitrary, the choice turns out to be very useful when the
turbulent flow is chaperoned by a few invariant solutions and their
unstable manifolds, as for example in low Reynolds number \pCf\
\citep{GHCW07}. Such visualisations are essential
to uncovering the interrelations between
invariant solutions, and constructing symbolic dynamics partitions of
\statesp\ needed for a systematic exploration of turbulent
dynamics. This is the key challenge we address here for the case of
turbulent pipe flows.

\subsection{Symmetries of pipe flow}
\label{s:SymmPipe}

In many physical applications equations such as those of Navier--Stokes
retain their form under symmetry transformations. Consider the
Navier--Stokes equations in the \statesp\ formulation \refeq{symbolicNS}.
A flow $\dot{\ssp}= \vel(\ssp)$ is said to be $\Group$-\emph{equivariant}
if the form of evolution equations is left invariant by the set of
transformations $\LieEl$ that form the group of symmetries of the
dynamics $\Group$,
\beq
\vel(\ssp)=\LieEl^{-1} \, \vel(\LieEl \, \ssp)
\,,\qquad \mbox{for all } \LieEl \in {\Group}
\,.
\ee{eq:FiniteRot}
On an infinite domain and in the absence of boundary conditions, the
\NSe\ are equivariant under translations, rotations, and $\bx \to -\bx$,
$\bu \to -\bu$ inversion through the origin \citep{frisch}. In pipe flow
the cylindrical wall restricts the rotation symmetry to rotation about
the $z$-axis, and translations along it. Let $\LieEl(\gSpace,\shift)$ be
the shift operator such that $\LieEl(\gSpace,0)$ denotes an azimuthal
rotation by $\gSpace$ about the pipe axis, and $\LieEl(0,\shift)$ denotes
the stream-wise translation by $\shift$; let $\sigma$ denote reflection
about the $\theta=0$ azimuthal angle:
\bea
\LieEl(\gSpace,\shift) \, [u,v,w,p](r,\theta,z)
        & = & [u,v,w,p](r,\theta-\gSpace,z-\shift)
			  \continue
\sigma \, [u,v,w,p](r,\theta,z) \;\; & = & [u,-v,w,p](r,-\theta,z)
\,.
\label{pipeSymms}
\eea
The Navier--Stokes equations for pipe flow are equivariant under these
transformations. The symmetry group of stream-wise periodic pipe flow is
thus $\Group = \On{2}_\theta \times \SOn{2}_z = \Dn{1} \ltimes
\SOn{2}_{\theta} \times \SOn{2}_z$, where $\Dn{1} = \{ e,\, \sigma \}$
denotes azimuthal reflection, $\ltimes$ stands for a semi-direct product
(in general, reflections and rotations do not commute), and the
subscripts $z,\theta$ indicate stream-wise translation and azimuthal
rotation respectively. For an assessment of the discrete symmetries in
pipe flow see \refappe{appe:DiscSymmPipe}.

Whilst the flow equations are invariant under $\Group$, the state of flow
typically is not. Only the laminar Hagen--Poiseuille \eqv\ is invariant
under all of $\Group$, whereas a generic turbulent state has only the
trivial symmetry group $\{e\}$. In this paper we restrict our
investigations to dynamics restricted to the `shift-and-reflect' symmetry
subspace \refeq{ShiftRefl},
\beq
   S = \{e,\sigma\LieEl_z\},
\ee{ShiftReflOnl}
\ie\ velocity fields \refeq{pipeSymms} that satisfy
\(
[u,v,w,p](r,\theta,z) =
[u,-v,w,p](r,-\theta,z-L/2)
\,.
\)
In addition, in some of the simulations (e.g.\ figure~\ref{fig:M1loc2})
we further impose the `rotate-and-reflect' symmetry
\beq
   Z_2 \ = \{e,\sigma\LieEl_\theta\},
\ee{ShiftRot}
which is possessed by the highly symmetric waves found by
\citet{Pringle09}. In this case the velocity field also
satisfies
\(
[u,v,w,p](r,\theta,z) =
[u,-v,w,p](r,\pi/2-\theta,z)
\,.
\)

It is worth emphasising that by imposing the symmetry $S$, rotations
are prohibited and hence we consider only the simplest example of a
continuous group, the stream-wise one-parameter rotation group
$\SOn{2}_z$, omitting the subscript $z$ whenever that leads to no
confusion. In the literature (see, e.g.\ \cite{ReSaTkYa11}) such
\SOn{2} is often referred to as the circle group $S^1$.

\subsection{Symmetry-induced coordinate frames}
\label{s:symm}

So far we have not offered any advice as to the choice of basis
vectors in constructing \statesp\ coordinates \refeq{intrSspTraj}. In
this section we show that the presence of a continuous symmetry
suggests two natural mutually orthogonal basis vectors, the group
action tangent and curvature vectors,
suitable to local visualisations of group orbits.

Consider the one-parameter rotation group \SOn{2} acting on a smooth
periodic function $u(\theta + 2\pi) = u(\theta)$ defined on the
domain $\theta \in [0,2\pi)$, expanded in the Fourier basis
\[ 
   u(\theta) = \sum \ssp_m \mathrm{e}^{\mathrm{i}m\theta}.
\] 
Here $u$ is real, so $\ssp_m=\ssp_{-m}^*$. Let us parametrise forward
translations by the continuous parameter $\gSpace$,
\(
    \LieEl(\gSpace)\,u(\theta) = u(\theta-\gSpace)
\,,
\)
or, in  Fourier space,
\(
   \LieEl(\gSpace) \,\ssp = \mathrm{diag}\{ \mathrm{e}^{-\mathrm{i}m\gSpace} \} \,\ssp
\,.
\)
The tangent to the group orbit at point $\ssp$ is then given by
the first derivative with respect to the group parameter,
and the direction of curvature by the second derivative,
\bea
   {\bf t}(\ssp) ~~~ &=& \;
   \lim_{\gSpace\to 0}
   \left(\LieEl(\gSpace)\,\ssp - \ssp\right)/\gSpace
   = \mathrm{diag}\{ -\mathrm{i}m \} \, \ssp = \Lg \ssp,
\label{eq:tang}\\
   \kappa(\ssp)\, {\bf n}(\ssp) \;&=&\; \Lg^2 \ssp  = - \mathrm{diag}\{m^2\} \, \ssp
   \,,
\label{eq:curv}
\eea
where $\normVec$ is a unit vector normal to the tangent and
$1/\kappa$ is the radius of curvature. The pair of unit vectors
\beq
\{{\be_n},{\be_{n+1}}\} =
\{\groupTan(\ssp)/\Norm{\groupTan(\ssp)},\normVec(\ssp)\}
\ee{FrenetFrame}
forms a local orthogonal Frenet--Serret frame at \statesp\
point $\ssp$, and can be useful
in constructing the \statesp\ basis vector set \refeq{intrSspTraj}. For
example, in \reffig{f:MeanVelocityFrame} the \rpo\ $\RPO{36.72}$ is
projected onto the $3$\dmn\ orthogonal frame
\beq
\{{\be}_1,{\be}_2,{\be}_3\}
=
    \left\{
{\groupTan(\slicep)}/{\Norm{\groupTan(\slicep)}},
\,\normVec(\slicep),\,
(\sspRed_d-\slicep)_\perp/\Norm{\sspRed_d-\slicep}_\perp
    \right\}
\ee{FrenetFrame1}
where $\slicep=\ssp(0)$ is a point on the \rpo\ (such fluid snapshot is
called `template' or `reference state' in what follows), $\sspRed_d$ is
the most distant point  from $\sspRed$ along its symmetry-\reducedsp\
\po\ $\sspRed(\zeit)$, measured in the energy norm \refeq{innerproduct},
and
$(\sspRed_d-\slicep)_\perp$ is the component of their separation vector,
Gram--Schmidt orthogonalised to $\{\be_1,\be_2\}$.

In what follows we consider time-dependent group parameters
$\gSpace(\zeit)$, and the associated \emph{\phaseVel} $\dot{\gSpace}$ along
the group tangent evaluated at the \statesp\ point $\ssp$ is given by
\beq
\LieEl^{-1}\dot{\LieEl} \,\ssp
     =\mathrm{e}^{-\gSpace \Lg} \,
\left(\frac{\mathrm{d} ~~}{\mathrm{d} \, \zeit} \, 
                             \mathrm{e}^{\gSpace \Lg}\right)\ssp
     =\dot{\gSpace}\cdot \groupTan(\ssp)
\,.
\ee{CartanDer}
This formula for the {\phaseVel} is known as the `Cartan derivative';
for $N$-parameter continuous symmetry the dot product is $N$-dimensional,
as in \refeq{phaseVel}.

\subsection{Relative invariant solutions}
\label{s:RelInvSol}

In systems with continuous symmetries there are important classes of
invariant solutions referred to as `relative' or `equivariant'
\citep{Huyg1673,Poinc1896}. In pipe flows one expects to find
\reqva\ and \rpo s \citep{Rand82} associated with the translational
and rotational symmetries of the flow. Although these unstable
flow-invariant solutions can only be computed numerically, they are
`exact' in the sense that they converge to solutions of the \NS\
equations as the numerical resolution increases.

A {\em relative equilibrium} (labelled here $\REQV{}{}$ for
travelling wave) is a dynamical
orbit whose velocity field \refeq{symbolicNS} lies within the group
tangent space
\beq
\vel(\ssp)  =  \velRel \cdot \groupTan(\ssp)
\,,
\ee{phaseVel}
with a constant {\phaseVel}
$(\dot{\gSpace}_1,\cdots,\dot{\gSpace}_N)=(\velRel_1,\cdots,\velRel_N)$
and
$\dot{\gSpace}$ defined in \refeq{CartanDer}.
Here $N$ is the
dimension of the continuous symmetry. In pipe flow $N=2$ and
$\{\gSpace_n\}=\{\gSpace,\shift\}$, corresponding to rotations and
translations. For a \reqv, time evolution is
confined to the group orbit
\beq
\ssp(\zeit)  =  \LieEl(\velRel\zeit) \, \ssp(0)
\,,\qquad
\ssp(\zeit) \in \pS_{\REQV{}{}}
\,.
\ee{phaseVel1}

As a travelling wave explores only its group orbit,
a \reqv\ is
\emph{not} a \po. Rather, as all states in a group orbit are physically
the same state, this is a generalised \eqv.  In pipe flow relative
equilibria can propagate in the stream-wise direction $z$ (travelling
waves), in azimuthal $\theta$ direction (rotating waves), or both.
However, in the shift-and-reflect subspace \eqref{ShiftReflOnl}
considered here, rotations are precluded. In this case only stream-wise
\reqva\ are permitted, satisfying \refeq{phaseVel1}
\beq
   \vec{f}(\vec{u}(0),\zeit)
\,=\, \LieEl(0,-c\zeit)\,\vec{u}(\zeit) - \vec{u}(0) = \vec{0}
\,,
\ee{pipeAxTW}
where $c$ is the stream-wise {\phaseVel}.

A {\rpo} $p$ is an orbit in {\statesp} $\pS$ which exactly recurs
\[ 
\ssp(\zeit) = \LieEl_p \, \ssp(\zeit + \period{p} )
    \,,\qquad
\ssp(\zeit) \in \pS_p
\] 
after a fixed {relative period} $\period{}$, but shifted by a fixed group
action ${\LieEl}$ that maps the endpoint $\ssp (\period{} ) $ back into
the initial point cycle point $\ssp (0) $. In pipe flow, a \rpo\ $p$ is a
time-dependent velocity field
\beq
\bu_p(r,\theta,z,\zeit)=\bu_p(r,\theta+\gSpace_p,z+ \shift_p,\zeit+\period{p})
\ee{RPOpipe}
that recurs after time $\period{p}$, rotated and shifted by $\gSpace_p$
and $\shift_p$. In our Newton search for a \rpo\ $p$, we seek the zeros
of
\beq
   \vec{f}(\vec{u}(0),\period{},\shift)
\,=\, \LieEl(0,-\shift)\,\vec{u}(\period{}) - \vec{u}(0) \,=\, \vec{0}
\,,
\ee{eq:fRPO}
starting with a guess for the initial state of fluid
$\vec{u}$, period $\period{}$, and shift $\shift$.

Continuous symmetry parameters (`phases' or `shifts')
$\{\gSpace_n\}=\{\gSpace_p,\shift_p\}$ are real numbers, so ratios
$\pi/\gSpace_n$ are almost never rational, and \rpo s are almost never
periodic. In pipe flow the time evolution of a \rpo\
sweeps out quasi-periodically the $3$\dmn\ group orbit $\pS_p$ without
ever closing into a \po.

\section{Reduction of continuous symmetry}
\label{s:slice}

We have seen that in presence of the continuous $\SOn{2}$ symmetry,
\reqva\ and \rpo s are 2- and 3-dimensional manifolds of physically
equivalent states generated by axial and azimuthal shifts.
How are we to compare a pair of such states?
We start by determining the minimal distance between them.

\begin{figure}
 \begin{center}
  \setlength{\unitlength}{0.40\textwidth}
(a)~~
  \begin{picture}(1,1.07471658)%
    \put(0,0){\includegraphics[width=\unitlength]{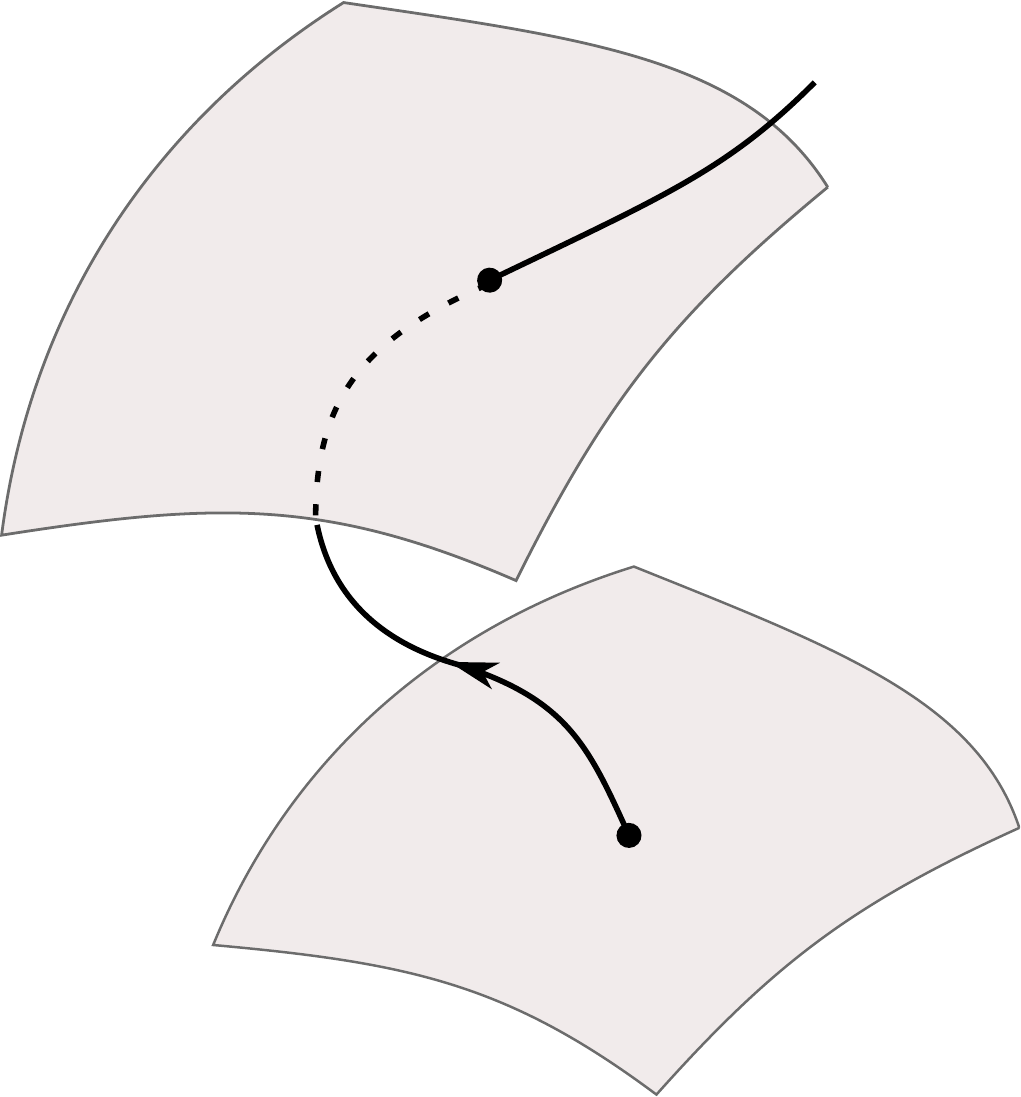}}%
    \put(0.28879298,1.02196543){\color[rgb]{0,0,0}\rotatebox{-22.37140782}{\makebox(0,0)[lb]{\smash{$\pS_{\ssp(\zeit)}$}}}}%
    \put(0.55566402,0.45078735){\color[rgb]{0,0,0}\rotatebox{-16.6673442}{\makebox(0,0)[lb]{\smash{$\pS_{\ssp(0)}$}}}}%
    \put(0.63028127,0.18433597){\color[rgb]{0,0,0}\rotatebox{0.03136739}{\makebox(0,0)[lb]{\smash{$\ssp(0)$}}}}%
    \put(0.46253394,0.70182304){\color[rgb]{0,0,0}\rotatebox{0.03136739}{\makebox(0,0)[lb]{\smash{$\ssp(\zeit)$}}}}%
    \put(0.03852492,0.09250899){\color[rgb]{0,0,0}\rotatebox{0.11031334}{\makebox(0,0)[lb]{\smash{$\pS$}}}}%
  \end{picture}%
~~(b)
  \begin{picture}(1,1.07315413)%
    \put(0,0){\includegraphics[width=\unitlength]{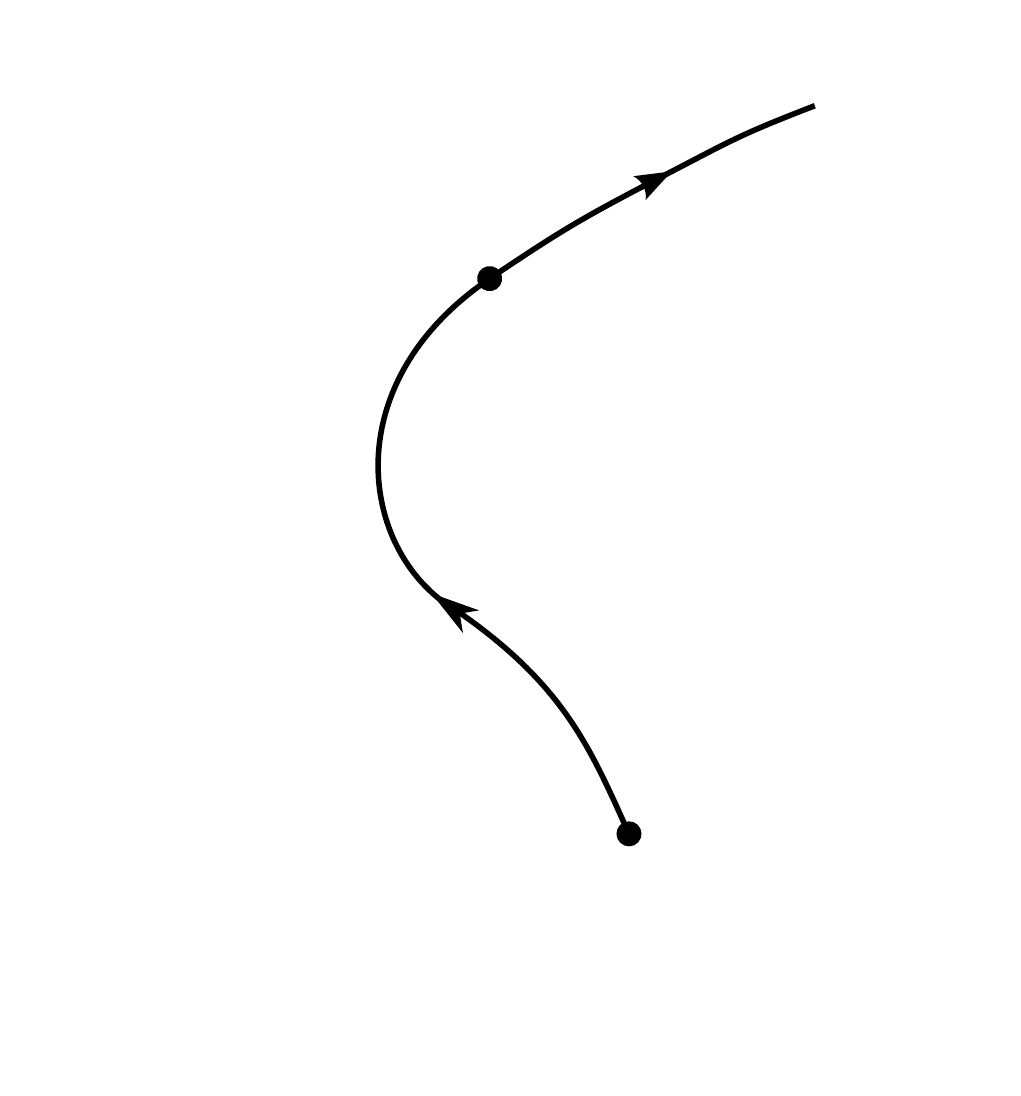}}%
    \put(0.19912369,0.17144733){\color[rgb]{0,0,0}\rotatebox{0.11031334}{\makebox(0,0)[lb]{\smash{$\pSRed$}}}}%
    \put(0.63028127,0.18433598){\color[rgb]{0,0,0}\rotatebox{0.03136739}{\makebox(0,0)[lb]{\smash{$\sspRed(0)$}}}}%
    \put(0.46253394,0.70182305){\color[rgb]{0,0,0}\rotatebox{0.03136739}{\makebox(0,0)[lb]{\smash{$\sspRed(\zeit)$}}}}%
  \end{picture}%
 \end{center}
  \caption{\label{fig:BeThTraj}
(a)
The group orbit $\pS_{\ssp(0)}$ of \statesp\ point $\ssp(0)$, and the
group orbit $\pS_{\ssp(\zeit)}$ reached by the trajectory $\ssp(\zeit)$
time $t$ later.
(b)
Symmetry reduction $\pS \to \pSRed$ replaces each full \statesp\ group
orbit $\pS_{\ssp}\subset\pS$ by a single point in the \reducedsp\
$\sspRed \in \pSRed$.
  }
\end{figure}

The \emph{group orbit} $\pS_\ssp $ of a \statesp\ point $\ssp \in \pS$ is
traced out by the set of all group actions
\beq
\pS_\ssp = \{\LieEl\,\ssp \mid \LieEl \in {\Group}\}
\,.
\ee{sspOrbit}
Any state in the  group orbit set $\pS_{\ssp}$ is physically equivalent
to any other. The action of a symmetry group thus foliates the \statesp\
into a union of group orbits, \reffig{fig:BeThTraj}\,({\it a}).

\begin{figure}
  \centering
(a)\includegraphics[width=0.45\textwidth]{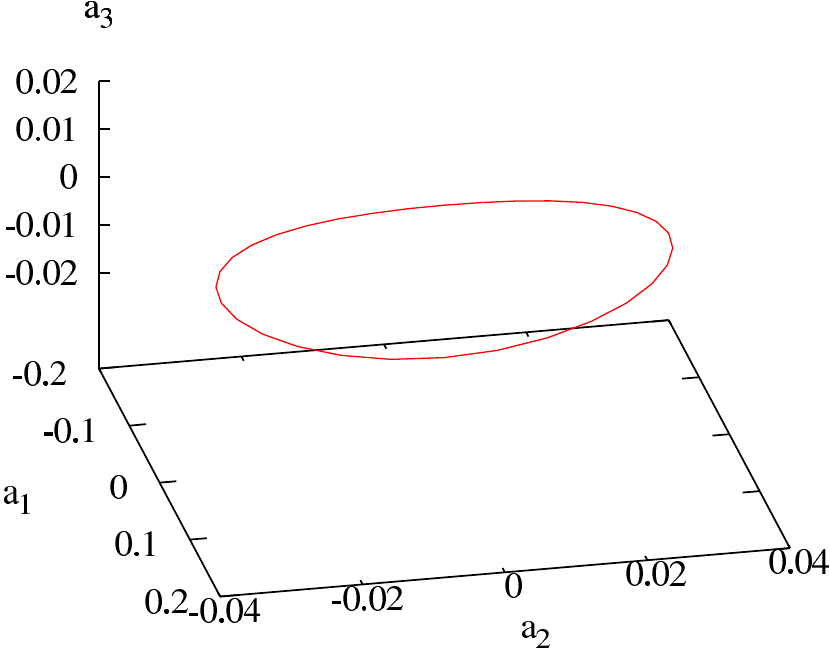}
(b)\includegraphics[width=0.45\textwidth]{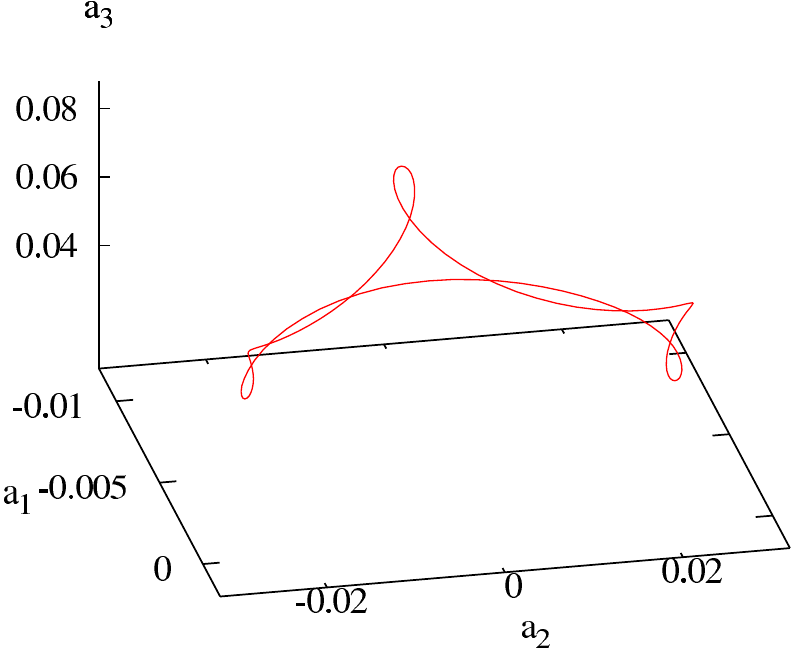}
  \caption{\label{fig:2840GOt135th0}
Projections of group orbits of two states $\ssp$ (in $\approx
100,000$-dimensional {\statesp}) onto stationary Frenet--Serret frames
given by unit vectors in the directions
$\{\groupTan_z(\slicep),\groupTan_\theta(\slicep),\normVec_z(\slicep)\}$,
see \refeq{FrenetFrame}. The state in (a) is a very smooth state, the
`lower-branch' travelling wave LB, whereas in (b) it is a snapshot from a
turbulent run. The group orbits are generated by all possible axial
shifts $\LieEl(0,\shift)\,\ssp$, and plotted relative to a template point
$\slicep$. In (a) the state $\ssp = \slicep=\ssp_\mathrm{LB}(0)$ belongs
to the ``lower-branch'' \reqv\ $\pS_\mathrm{LB}$ described in
\refsect{s:eqbSols}; in (b) $\ssp$ is a `typical' turbulent state
shapshot with its group orbit as seen from the template
$\slicep=\ssp_\mathrm{ML}$. Group orbits are only topologically circles;
for strongly nonlinear, turbulent states many Fourier modes are of
comparable magnitude, with their sums resulting in highly convoluted
group orbits such as (b).
  }
\end{figure}

\begin{figure}
  \centering
(a)\includegraphics[width=0.45\textwidth]{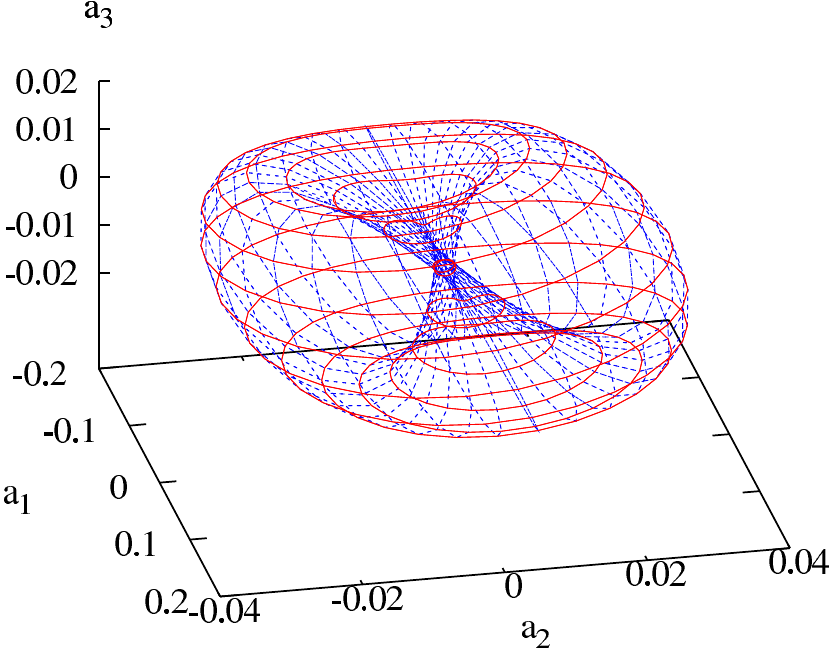}
(b)\includegraphics[width=0.45\textwidth]{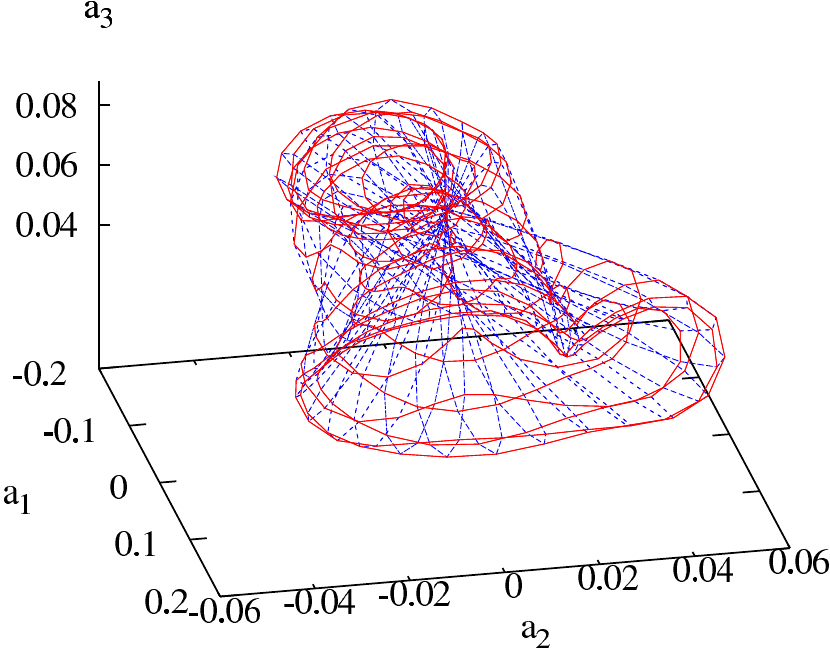}
  \caption{\label{fig:2830GO6}
As \reffig{fig:2840GOt135th0}, but with the
full 2\dmn\ $\SOn{2}_{\theta} \times \SOn{2}_z$ group orbits traced out
by shifts in both $z$ and $\theta$. Loops in solid red correspond to
shifts in $z$, dashed blue loops to shifts in $\theta$.
  }
\end{figure}

For the example at hand, a pipe flow (or a \pCf) with two periodic
boundary conditions, the symmetry group $\Gpipe$ contains two commuting
\SOn{2} rotations. Each \SOn{2} subgroup group orbit is (topologically) a
circle, see \reffig{fig:2840GOt135th0}, and together they sweep out a
$T^2$ torus, see \reffig{fig:2830GO6}.

The goal of \emph{symmetry reduction} is to replace each group orbit by a
unique point in a lower-dimensional symmetry-\reducedsp\ $\pSRed =
\pS/\Group$, as sketched in \reffig{fig:BeThTraj}. Several symmetry
reduction schemes are reviewed in \refref{SiCvi10}. Here we shall
describe the \mslices\ \citep{rowley_reconstruction_2000,BeTh04,FrCv11}, the
only method that we find practical for a symmetry reduction of
turbulent solutions of highly nonlinear flows, see \refsect{s:rpos}.

In the \mslices\ the symmetry reduction is achieved by cutting the group
orbits with a finite set of hyperplanes, one for each continuous group
parameter, with each group orbit of symmetry-equivalent points
represented by a single point, its intersection with the \slice. The
procedure is akin to (but distinct from) cutting across continuous-time
parametrised trajectories by means of Poincar\'e sections. As is the case
for Poincar\'e sections, choosing a `good' \slice\ is a dark art. Our
guiding principle is to chose a \slice\ such that the distance between a
`{\template}' state {\slicep} and nearby group orbits is
\emph{minimised}, \ie, identify the point $\sspRed$ on the group orbit
\refeq{sspOrbit} of a nearby state $\ssp$ which is the closest match to
the {\template} point {\slicep}.

\subsection{\Mslices; local charts}

After some experimentation and observations of turbulence in a given
flow, one can identify a set of dynamically important unstable
{\recurrStr s}.  For example, coherent streaky structures have been
observed in pipe flow at transitional $\Reynolds$ \citep{science04} through to
very high $\Reynolds$ \citep{kim99} where `very large scale motions' have
length scales comparable to the pipe radius.  Streaky structures are also
observed in the buffer layer of turbulent flows with a characteristic
span-wise wavelength of approximately 100 wall units \citep{KRSR67}.

We shall refer to this catalogue of $n$ representative snapshots
or `reference states', either precomputed or
experimentally measured, as  \emph{\template s}
\citep{rowley_reconstruction_2000}, each an instantaneous state of the
$3D$ fluid flow represented by a
\emph{point} $\slicep{}^{(j)}$, $j=1,2,\cdots,n$, in the
\statesp\ $\pS$ of the system.
Symmetries of the flow (i.e.\ the
$\LieEl\in\Group$) are then used to shift and rotate the {\template}
$\slicep$ until it overlies, as well as possible, the {\cohStr} of
interest $\ssp$, by minimising the distance
\beq
\Norm{\ssp - \LieEl(\gSpace)\,\slicep}
\, .
\ee{minDistance}
The entire group orbit of $\ssp$ is then replaced by the closest match to
the template pattern, given by $\sspRed=\LieEl^{-1}\ssp$, as shifting
does not affect the norm,
$\Norm{\ssp-\LieEl\,\slicep}=\Norm{\sspRed-\slicep}$.
The symmetry-\reducedsp\ $\pSRed$
(hereafter referred to as the `slice'),
of dimension $(d\!-\!1)$, consists of
the set of closest matches $\sspRed$, one element for each full \statesp\ $\pS$
group orbit; the hat on $\sspRed$
indicates the unique point on the group orbit of $\ssp$ closest to
the \template\ \slicep.

For the azimuthal $\SOn{2}_\theta$ rotations (and likewise for the
periodic pipe  $\SOn{2}_z$ stream-wise translations),
the minimal distance satisfies the extremum condition
\[
\frac{\partial}{\partial \gSpace} \Norm{\ssp - \LieEl(\gSpace)\,\slicep}^2
   =
2\, \braket{\ssp - \LieEl\,\slicep}{\Lg_\theta \,\LieEl\,\slicep}
   =
2\, \braket{\sspRed - \slicep}{\Lg_\theta \slicep}
   = 0
    \, ,
\]
given that group orbits are smooth differentiable manifolds. As
$\Norm{\LieEl(\gSpace)\slicep}$ is a constant, the group tangent vector
$\Lg_\theta \slicep$ evaluated at $\slicep$ \refeq{eq:tang} is normal to
$\slicep$, and the term $\braket{\slicep}{\Lg_\theta\,\slicep}$ vanishes
($\Lg_{\theta}$ is antisymmetric). Therefore the point $\sspRed$ on the
group orbit that lands in the \slice, satisfies the \emph{\slice\
condition}
\beq
\braket{\sspRed}{\sliceTan{\theta}} = 0
    \,,\quad
\sliceTan{\theta} = \Lg_\theta \slicep
    \,.
\ee{PCsectQ0}
The \slice\ so defined is thus a hyperplane that includes the origin,
normal to the \template\ group tangent evaluated at the \template.


 \begin{figure}
 \begin{center}
  \setlength{\unitlength}{0.40\textwidth}
(a)
  \begin{picture}(1,0.87085079)%
    \put(0,0){\includegraphics[width=\unitlength]{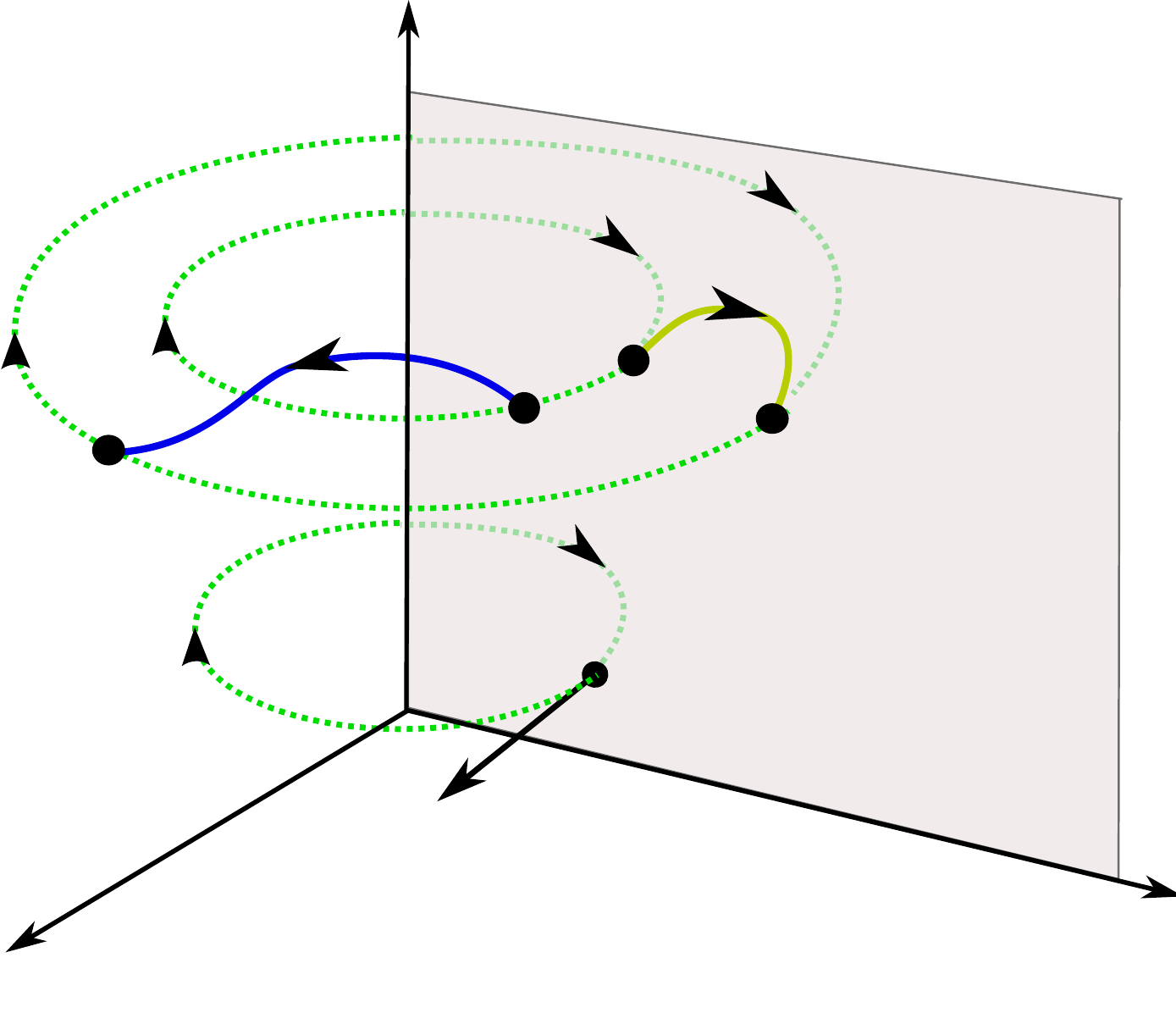}}%
    \put(0.82835155,0.19007659){\color[rgb]{0,0,0}\rotatebox{-14.84025424}{\makebox(0,0)[lb]{\smash{$\pSRed$}}}}%
    \put(0.07077338,0.28688228){\color[rgb]{0,0,0}\rotatebox{0.0313674}{\makebox(0,0)[lb]{\smash{$\LieEl\,\slicep$}}}}%
    \put(0.53023327,0.26593335){\color[rgb]{0,0,0}\rotatebox{0.0313674}{\makebox(0,0)[lb]{\smash{$\slicep$}}}}%
    \put(0.4284954,0.179285){\color[rgb]{0,0,0}\rotatebox{0.0313674}{\makebox(0,0)[lb]{\smash{$\sliceTan{}$}}}}%
    \put(0.00798985,0.42305068){\color[rgb]{0,0,0}\rotatebox{0.0313674}{\makebox(0,0)[lb]{\smash{$\ssp(\zeit)$}}}}%
    \put(0.65766235,0.45412105){\color[rgb]{0,0,0}\rotatebox{0.0313674}{\makebox(0,0)[lb]{\smash{$\sspRed(\zeit)$}}}}%
    \put(0.06916446,0.74280851){\color[rgb]{0,0,0}\rotatebox{0.0313674}{\makebox(0,0)[lb]{\smash{$\LieEl(\zeit)$}}}}%
  \end{picture}%
~~~
(b)
  \begin{picture}(1,0.87085079)%
    \put(0,0){\includegraphics[width=\unitlength]{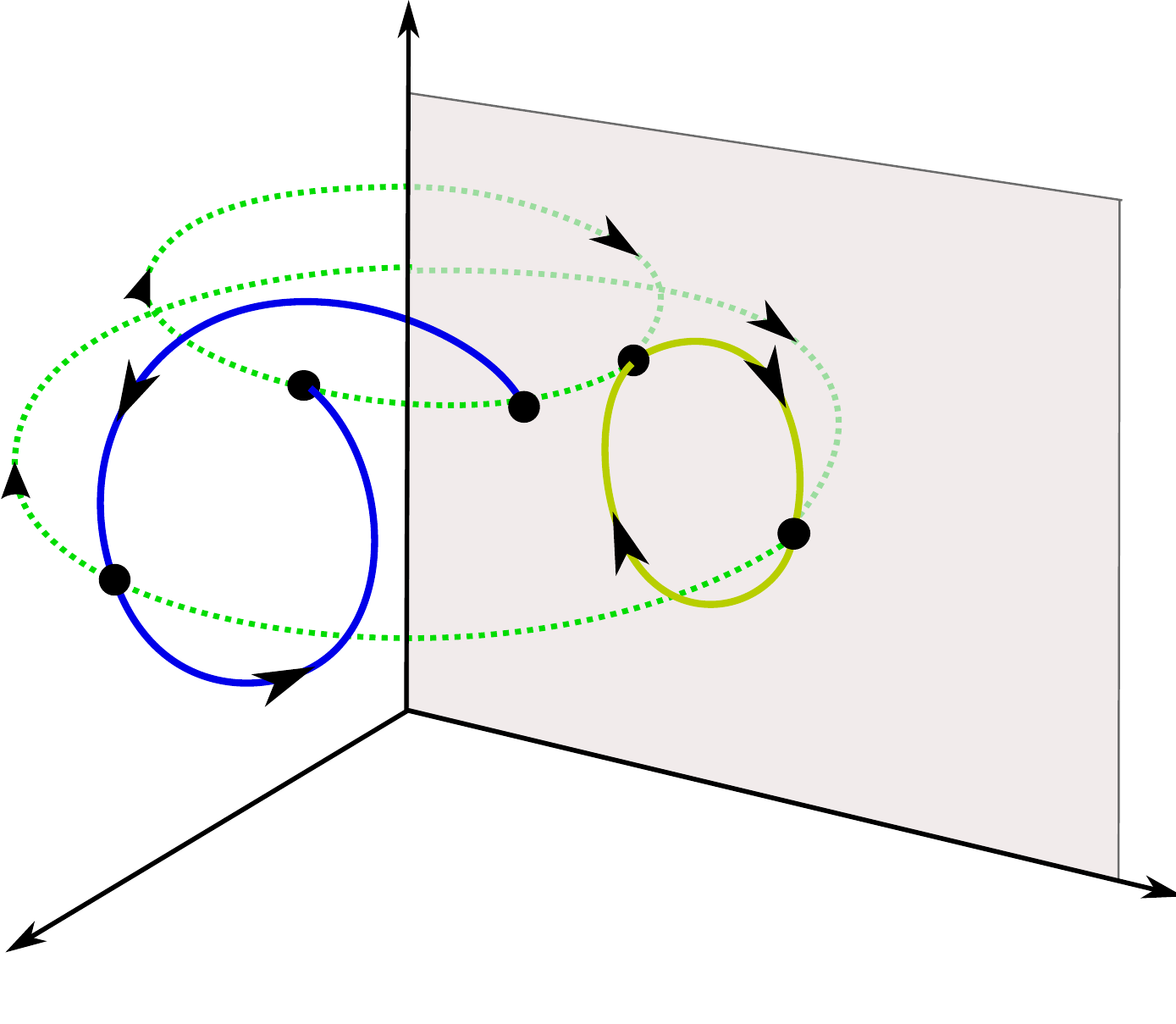}}%
    \put(0.82835153,0.19007656){\color[rgb]{0,0,0}\rotatebox{-14.84025432}{\makebox(0,0)[lb]{$\pSRed$}}}%
    \put(0.40925459,0.45713857){\color[rgb]{0,0,0}\rotatebox{0.0313674}{\makebox(0,0)[lb]{\smash{$\ssp(0)$}}}}%
    \put(0.71354118,0.39765314){\color[rgb]{0,0,0}\rotatebox{0.0313674}{\makebox(0,0)[lb]{\smash{$\sspRed(\zeit)$}}}}%
    \put(0.13171187,0.38813817){\color[rgb]{0,0,0}\rotatebox{0.0313674}{\makebox(0,0)[lb]{\smash{$\LieEl(\zeit)$}}}}%
    \put(0.02168739,0.31359574){\color[rgb]{0,0,0}\rotatebox{0.0313674}{\makebox(0,0)[lb]{\smash{$\ssp(\zeit)$}}}}%
    \put(0.15576193,0.48769256){\color[rgb]{0,0,0}\rotatebox{0.0313674}{\makebox(0,0)[lb]{\smash{$\ssp(\period{})$}}}}%
    \put(0.54113911,0.50476963){\color[rgb]{0,0,0}\rotatebox{0.0313674}{\makebox(0,0)[lb]{\smash{$\sspRed(0)$}}}}%
  \end{picture}%
 \end{center}
 \caption{\label{fig:slice}
The \mslices, a \statesp\ visualisation:
(a)
\Slice\ $\pSRed \supset \pS/\Group$ lies in the $(d\!-\!N)$\dmn\
hyperplane \refeq{PCsectQ0} normal to $\sliceTan{}$, where $\sliceTan{j}$
span the  $N$\dmn\ space tangent to the group orbit $\LieEl\,\slicep$
(dotted line) evaluated at the {\template} point $\slicep$. The
hyperplane intersects {\em all} full \statesp\ group orbits (green
dashes). The full \statesp\ trajectory $\ssp(\zeit)$ (blue) and the
\reducedsp\ trajectory $\sspRed(\zeit)$ (green) are equivalent up to a
`moving frame' rotation $\ssp(\zeit)=\LieEl(\zeit)\,\sspRed(\zeit)$, where
$\LieEl(\zeit)$ is a shorthand for $\LieEl(\gSpace(\zeit))$.
(b)
In the full \statesp\ $\pS$ a \rpo\ $\ssp(0) \to \ssp(\zeit) \to
\ssp(\period{})$ returns to the group orbit of $\ssp(0)$ after time
$\period{}$ and a rotation by $\LieEl$,  $\ssp(0)=\LieEl \, \ssp
(\period{})$. For flows with continuous symmetry a generic \rpo\ fills
out quasi-\-periodically what is topologically a torus. In the \slice\
$\pSRed$ the symmetry-reduced orbit is periodic, $\sspRed(0) =
\sspRed(\period{})$. This is a highly idealised sketch: A group orbit is
a $N$\dmn\ manifold, and even for $\SOn{2}$ it is usually only
topologically a circle (see \reffig{fig:2840GOt135th0}), and can
intersect a hyperplane any number of times  (see
\reffig{fig:sliceimage}{\it a}).
 }
 \end{figure}

When $\ssp$ is varies in time, $\dot{\ssp}=\vel(\ssp)$,
the template $\slicep$ tracks the motion
using the slice condition \refeq{PCsectQ0} to
minimise $\Norm{\ssp(\zeit)-\LieEl(\gSpace(\zeit))\slicep}$,
and the
full-space trajectory $\ssp(\zeit)$ is thus rotated into the {\reducedsp},
$\sspRed(\zeit) = \LieEl^{-1}\,\ssp(\zeit)$,
by appropriate
\emph{moving frame} \citep{CartanMF,FelsOlver98,FelsOlver99,OlverInv}
angles $\{\gSpace(\zeit)_n\}$, as depicted in
\reffig{fig:slice}\,({\it a}).
Specializing to $\SOn{2}$, one can write the equations for the
\reducedsp\ flow, $\sspRed(\zeit) \in \pSRed$ confined to the \slice,
$\dot{\sspRed} = \velRed(\sspRed)$, as
\bea
\velRed(\sspRed) &=& \vel(\sspRed)
     \,-\, \dot{\gSpace}(\sspRed) \, \groupTan(\sspRed)
\label{EqMotMFrame}\\
\dot{\gSpace}(\sspRed) &=& \braket{\vel(\sspRed)}{\sliceTan{}}
                       /\braket{\groupTan(\sspRed)}{\sliceTan{}}
\,.
\label{reconstrEq}
\eea
In other words, $\vel$, the velocity in the full \statesp, can be written
as the sum of $\velRed$, the velocity component in the \slice, and
$\dot{\gSpace}\,\groupTan$, the Cartan derivative \refeq{CartanDer} or
the velocity component
along the group tangent directions.
The $\dot{\gSpace}$ equation is the {\em reconstruction equation}: its
integral keeps track of the group shifts in the full \statesp. In
particular, if  \sspRed\ is a point on a \reqv\ \refeq{phaseVel}, the
full \statesp\ velocity equals the {\phaseVel}, and $\velRed(\sspRed) =
0$, \ie, \reqva\ are always reduced to \eqva\ in the \slice. It should be
emphasised that we never integrate the reduced equations
\refeq{EqMotMFrame}; numerical simulations are always carried out in the
full \statesp. Slicing is implemented as postprocessing of numerical or
experimental data, by rotating full \statesp\ trajectories into the
\slice, as in \reffig{fig:slice}.

\subsection{Charting the \reducedsp; a global atlas}
\label{s:chartingslice}

\begin{figure}
   \centering
(a)\includegraphics[width=0.45\textwidth]{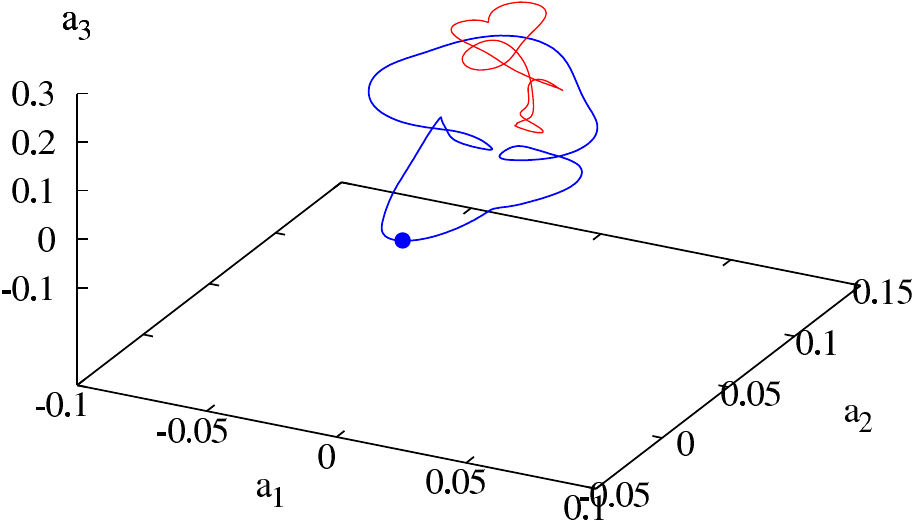}
(b)\includegraphics[width=0.45\textwidth]{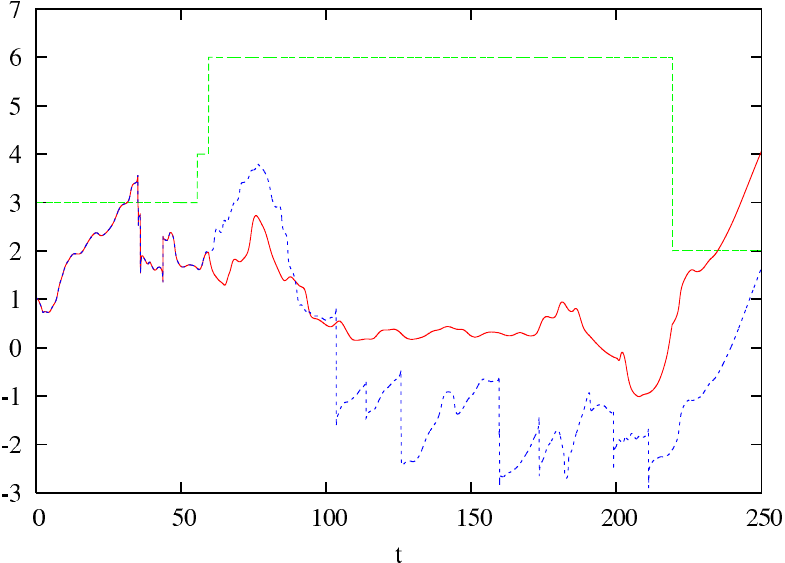}
   \caption{\label{fig:sliceimage}
   (a)
      Every slice hyperplane cuts every group orbit at least twice  (see
      \reffig{fig:slice}). An
      $\SOn{2}$ \rpo\ is topologically a torus, so the two cuts are the
      two \po\ images of the same \rpo, the good close one, and the bad
      distant one, on the other side of {\sliceBord}, and thus not in the
      slice. Here this is illustrated by close cut (blue, solid line) of
      the \rpo\ $\RPO{36.72}$ torus, \reffig{f:MeanVelocityFrame}\,({\it
      b}), plotted together with the most distant cut (red, dashed), in
      the same slice hyperplane, but not in the slice.
   (b)
      Comparison of symmetry-reduced trajectory using a single
      template ML (blue, short-dash) with the same trajectory symmetry
      reduced using the \reqva\ in \reftab{tab:RPOs} as template (red,
      solid) indexed by $j(\zeit)$ ($j = 1$ laminar; 2 LB; 3 ML; 4 MU;
      5 UB; 6 S2U; 7 S2L), with template at time $\zeit$ indicated on
      left ordinate by (green, long-dash) line. Right ordinate: the
      shift deviation from the mean shift,
      $\shift(\zeit)-\timeAver{\velRel}\,\zeit$, where
      $\timeAver{\velRel} \simeq 1.274$ is estimated by a long-time
      simulation. Both symmetry reductions begin with the same
      template and experience the same jumps in the shift starting at
      $\zeit\approx 40$. By starting to switch the templates at
      $\zeit\approx 60$, further jumps (seen for the blue, short-dash
      line) are avoided (red, solid line).
   }
\end{figure}

The \mslices\ as implemented here associates a \slice\
\refeq{PCsectQ0} to a \template. Our \slice\ is locally a hyperplane,
expected to be a good description of solutions similar to a given
template only in its neighbourhood. Nevertheless, as every group orbit has
a point closest to a given \template, and a \slice\ is the set of all
such group-orbit points, it slices the group orbits of \emph{all} full
\statesp\ points. The variational distance condition \refeq{PCsectQ0} is
an extremum condition, and as the group orbits of highly nonlinear states
are highly contorted (see \reffig{fig:2830GO6}{\it b}), the distance
function can have many extrema, and multiple sections by a \slice\
hyperplane. For example, a \rpo\ sweeps out a torus, and is always
intersected by a \slice\ hyperplane in two or
more \po\ sections,
once at the orbit's closest passage to the template, with positive curvature
\refeq{eq:curv},   and another time at the most distant passage, also
satisfying the slice condition \refeq{eq:slcond}, but with negative
curvature
(see \reffig{fig:sliceimage}{\it a}).

As explained in \cite{FrCv11}, a \slice\ hyperplane captures faithfully
neighboring group orbits as long as it \slice s them well; it does so
until it reaches the {\chartBord}, the set of points $\sspRSing$
sufficiently far from the \template, whose group orbits are grazed
tangentially rather than sliced transversely. For such grazing group
orbit the group tangent vector lies in the \slice, and is thus orthogonal
to the \slice\ tangent,
\beq
\braket{\groupTan(\sspRSing)}{\sliceTan{}}= 0
\, .
\ee{sspRSing}
The {\phaseVel} $\dot{\gSpace}(\sspRSing)$ in \refeq{reconstrEq} then
diverges. While such divergence is an avoidable nuisance, an artifact of
the symmetry reduction to a particular \slice\ hyperplane, it is a
numerical nuisance nevertheless.

For points beyond the {\chartBord} \refeq{sspRSing} the group orbits have
more than one intersection with the \slice. It is clear what
the trouble with any single \slice\ hyperplane is: the nonlinear flow of
interest is taking place on a highly contorted curved manifold embedded
in the $\infty$-dimensional \statesp, so a single template cannot be a
good match globally.
It is as good as a projection of the whole Earth onto a single flat map
centered on Ulan Bator.
The physical task is to, in order to chart the
\statesp\ of a turbulent flow, pick a set of qualitatively distinct
{\template s} $\slicep{}^{(j)}$ whose \slice s  $\pS{}^{(j)}$ span across
neighbourhoods of the qualitatively most important {\cohStr s}, and which
together capture all of the asymptotic dynamics and provide a global
atlas of the dimensionally \reducedsp\ $\pSRed = \pS/\Group$. The choice
of \template s should reflect the dynamically dominant patterns seen in
the solutions of nonlinear PDEs, one typical of, let us say, 2-roll
states, one for 4-roll states, and so on. Each \slice\ hyperplane comes
with its {\chartBord} hyperplane of points \sspRSing, defined by the
grazing condition \refeq{sspRSing}, beyond which it should not be
deployed as a chart. Together they `Voronoi' tessellate  the curved
manifold in which the symmetry-reduced strange attractor is embedded by a
finite set of hyperplane tiles.

For example, in reducing turbulent trajectories of \refsect{s:rpos}, we
deploy a set of \reqva\ as our templates. Each associated \slice\
$\pS{}^{(j)}$, provides a local chart at $\slicep{}^{(j)}$ for a
neighbourhood of an important, qualitatively distinct class of solutions.
In our simulations we keep checking the distance to the template of the
symmetry-reduced trajectory, and switch to the next template neighbourhood
before the trajectory encounters the {\chartBord} \refeq{sspRSing} of the
current one, as illustrated by \reffig{fig:sliceimage}\,({\it b}).

\section{How to slice a pipe}
\label{s:algorithm}

Slicing is independent of numerical representation.
We describe our implementation, however, using the
convenient discretisation for pipe flow of \refeq{pipeDiscr}.
The deviation velocity field $\vec{u}$ and deviation pressure in the
Navier--Stokes equations \refeq{NavStokesDev} are discretised as in
\refeq{pipeDiscr}, using Fourier modes in the axial and azimuthal
directions and a finite-differences in the radial direction, with
coefficients $\vec{u}_{nkm}$. The radial points, $r_n$ for
$n=1,2,\cdots,N$, are non-uniformly spaced, with higher resolution
towards the wall. Flow variables being real implies that the coefficients
satisfy $\vec{u}_{nkm}=\vec{u}_{n,-k,-m}^*$. Time-stepping has been
performed using a second-order predictor-corrector method with a
time-step of $\Delta t=0.0025$. To ensure dealiasing in the evaluation of
nonlinear terms, Fourier series are evaluated on $3K$ and $3M$ spatial
points in $z$ and $\theta$ respectively.  For the calculations presented,
a resolution of $(N,K,M)=(64,16,16)$ has been used, corresponding to
$64\times 48\times 48$ grid points.

\subsection{Rotation into the \slice}

In this paper we consider only shifts $\shift$ in the stream-wise
direction \refeq{pipeSymms}. Denoting our state by
$\ssp=(\vec{u}_{nkm})$, the group tangent
$\groupTan(\ssp) = \Lg_z\, \ssp$ to $\ssp$ in the direction of axial
shifts  is given by \refeq{eq:tang},
\beq
   \groupTan(\vec{u})_{nkm} = -2\alpha k\,\mathrm{i} \, \vec{u}_{nkm}
\,,
\eeq
and the shift $\shift(\zeit)$ of \statesp\ trajectory $\ssp(\zeit)$ into
the \slice\ is determined by the \slice\ condition \refeq{PCsectQ0},
\beq
   \label{eq:slcond}
   f(\shift(\zeit)) =
   \braket{\LieEl(0,-\shift(\zeit))\,{\ssp}(\zeit)}{{\sliceTan{}}} = 0
\,,
\eeq
where ${\sliceTan{}}$ is the group orbit tangent evaluated at a
\template\ state ${\slicep}$. As long as the norm is discretisation
independent, the \slice\ condition \refeq{eq:slcond} is independent of
the numerical representation of the flow $\vec{u}$, be it finite
difference, spectral, and so on. The slice condition is solved for
$\shift(\zeit)$ every few time steps using Newton's method, where a good initial
guess for $\shift(\zeit)$ is obtained from the previous value and
$\dot{\shift}(\zeit)$.

When ${\sspRed}(\zeit)$ is close to ${\slicep}$, the function $f(\shift)$
has only one root. When ${\ssp}(\zeit)$ is far from ${\slicep}$, however,
$f(\shift)$ may have many roots, pairs of which may disappear with time.
This would lead to a discontinuity in $\shift(\zeit)$.  As explained in
\refsect{s:chartingslice}, in order to avoid this, a global atlas has to
be pieced together from local \slice\ charts, fixed by a well-chosen set
of \template s $\slicep{}^{(j)}$ . Shifts $\shift_j(\zeit)$ are tracked
for each local \slice\ chart $\pS{}^{(j)}$, and the next \slice\
hyperplane $\pS{}^{(j+1)}$ with $\shift_{j+1}(\zeit)$ is selected whenever
the distance to the next {\template} minimises $\Norm{{\sspRed}(\zeit)-
\slicep{}^{(j+1)}}$.

\subsection{Dynamically important solutions and Newton's method}
\label{s:reqva}

For pipe flows many stream-wise \reqva\ satisfying
\refeq{pipeAxTW} are known, and can be used as the starting points for
our \rpo\ searches.
The most of known solutions have no azimuthal precession.  This is
usually imposed by symmetry, but one could argue that it is the strong
stream-wise advection that favours structures with very weak azimuthal
rotation speed, empirically $|\velRel_\theta| \leq O(10^{-3})$.
Stream-wise \reqva\ evolve in time along their group orbit, generated by
$\LieEl(0,\shift(\zeit))$. They therefore satisfy the \slice\ condition
\refeq{eq:slcond} for $\shift(\zeit)= \shift_0 + \velRel_z\zeit$.

The few pipe flow \rpo s that have been found prior to this study were
located via tracking a Hopf bifurcation off a {\reqv} solution
\citep{duguet08,mellibovsky11}.
These `modulated travelling waves,' here
referred to as `tiny' \reqva, stick close to their
mother orbits, and explore little of the \statesp, with temporal dynamics
barely distinguishable from parental \reqva. In contrast, in our Newton
searches for \rpo s, we seek the zeros of \rpo\ condition \refeq{eq:fRPO}
deep in the turbulent sea. The way in which the \mslices\ enables
one to find
initial guesses for $(\vec{u}(0),\period{},\shift)$, is the main
difference between this study and the previous searches for \rpo s in
pipe flows.

Here we take as initial guesses samples of nearly recurrent velocity
fields generated by long-time simulations of turbulent dynamics
\citep{pchaot,CviGib10}. The intent is to find the {\em dynamically most
important} solutions, by sampling the turbulent flow's natural measure.
In practice, sufficiently good full \statesp\ initial guesses for
$(\vec{u}(0),\period{},\shift)$ would be almost impossible to find.
Checking correlations between $\vec{u}(\zeit)$ and
$\LieEl(0,\shift)\,\vec{u}(\zeit-\period{})$ for each $\period{}$, and
more problematically, for all possible shifts $(\gSpace,\shift)$, is an
unrealistic task. The \mslices, however, enables us to determine close
recurrences from the symmetry-reduced time series, and locates the
dynamically most important solutions, \ie, those trajectories that are
most likely to be observed in a long-time turbulent simulation. The \rpo
s are reduced to \po s, whose unstable manifolds are much easier to track
in the \reducedsp. The \rpo\ shift $\shift$ is given by the
reconstruction equation, \refeq{reconstrEq}, or, in practice, by phase
shift $\shift(\period{})-\shift(0)$, where $\shift(t)$ is quickly
calculated by intermediate Newton steps.

With a good initial guess for $(\vec{u}(0),\period{},\shift)$, such a
system can be solved using a Newton scheme.  Two conditions in addition
to \refeq{eq:fRPO} need to be enforced: the Newton update should have no
component along the group orbit, $\braket{\vec{\delta
u}}{\groupTan(\vec{u})}=0$, and no component tangent to trajectory,
$\braket{\vec{\delta u}}{\dot{\vec{u}}}=0$. To solve this system a
`hookstep' trust-region variation to the Newton--Krylov method has been
implemented, similar to that of \cite{Visw07b}. This method greatly
increases the tolerance in the starting $(\vec{u}(0),\period{},\shift)$
required for convergence to an exact solution.

The radial and azimuthal components of the flow are typically smaller
than the stream-wise component by a factor of approximately $3$ to $10$.
The components $u$ and $v$, however, can be associated with `rolls' in
the flow that are as important to the self-sustaining mechanism of
turbulence as `streaks', associated with deviations in the $w$ component.
This observation motivates use of an empirical `compensatory norm',
\beq
   \Norm{\vec{u}}_c^2 \,=\, \braket{\vec{u}}{\vec{u}}_c
   \,=\, \frac{1}{2}\,\int_V (9\,u \cdot u + 9\,v \cdot v
                              + w \cdot w) \, \mathrm{d}V
\,,
\ee{compensNorm}
found to be useful for the calculation of recurrences and to assist
convergence in our Newton scheme.

\section{The sliced pipe}
\label{s:slicedWurst}

\begin{figure}
  \centering
  \begin{tabular}{ccc}
    LB & ML & MU \\
    \includegraphics[width=0.24\textwidth]{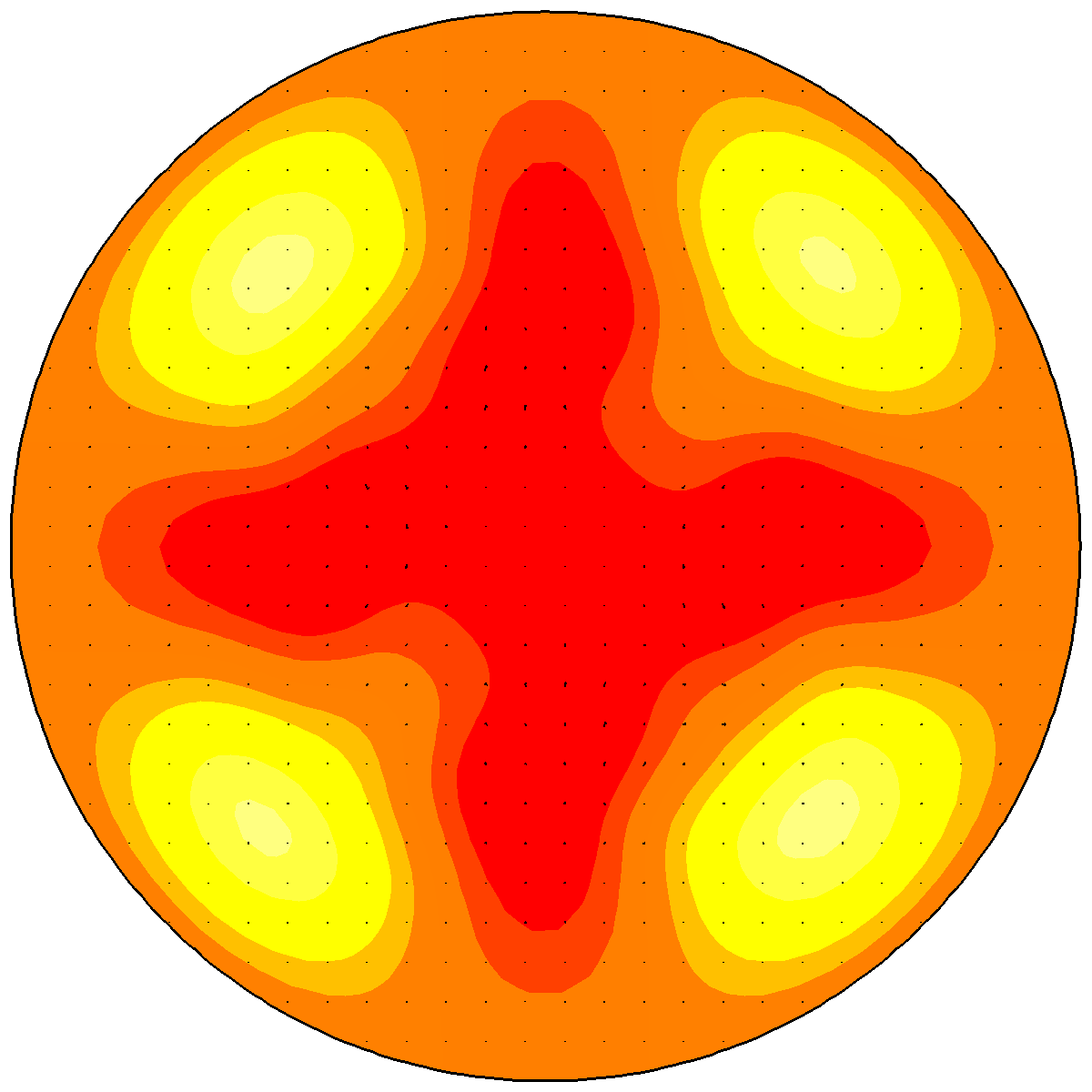}&
    \includegraphics[width=0.24\textwidth]{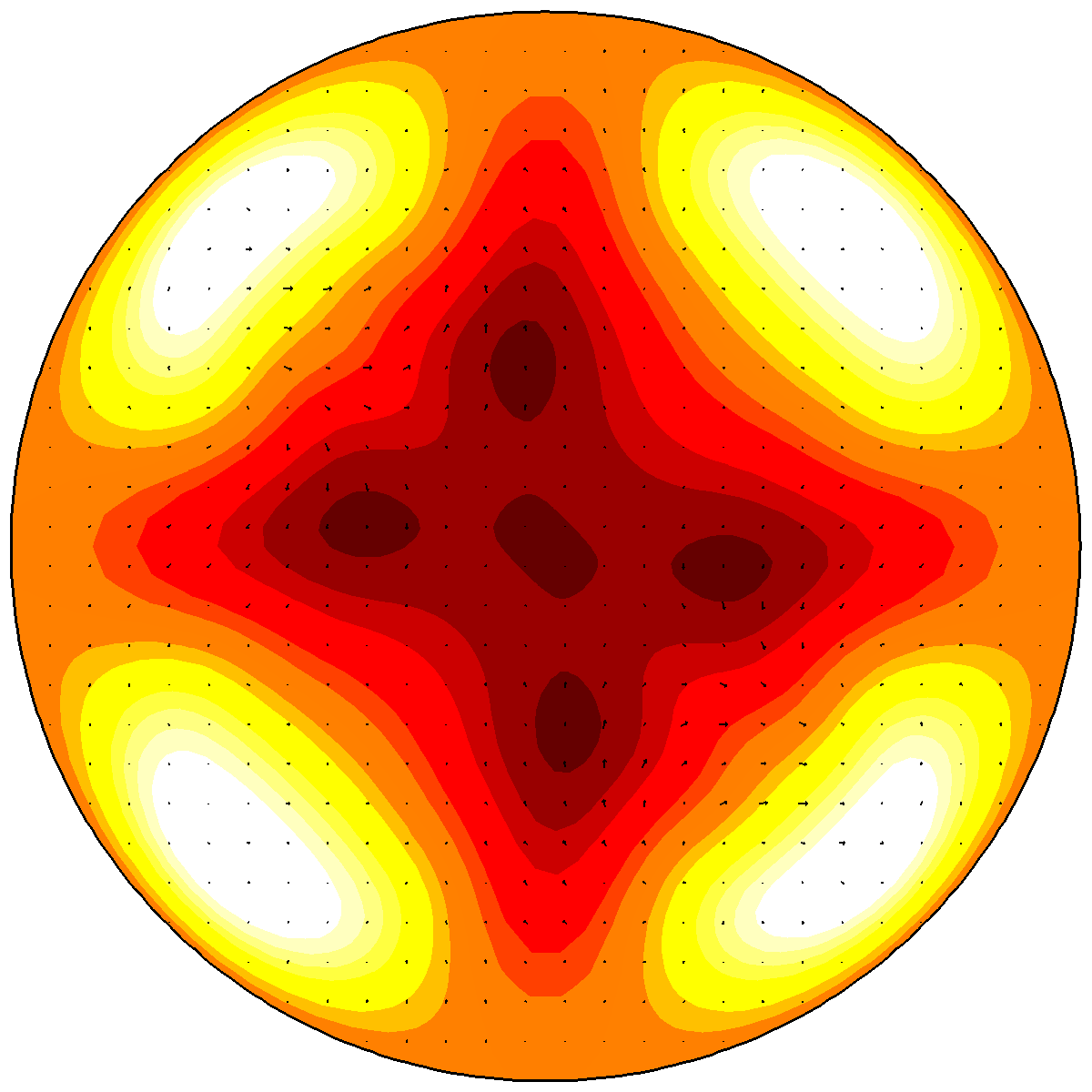}&
    \includegraphics[width=0.24\textwidth]{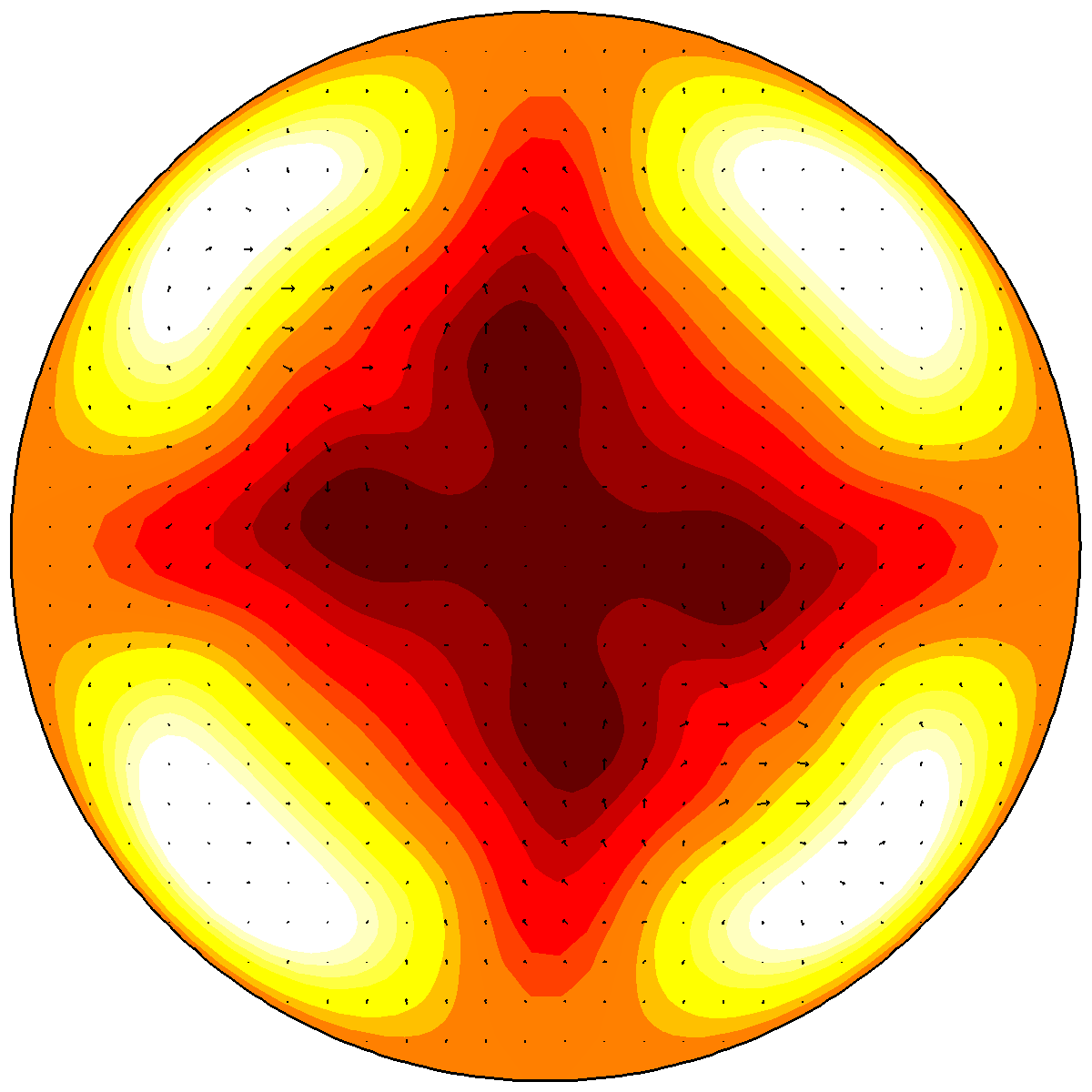}\\
    UB & S2U & S2L\\
    \includegraphics[width=0.24\textwidth]{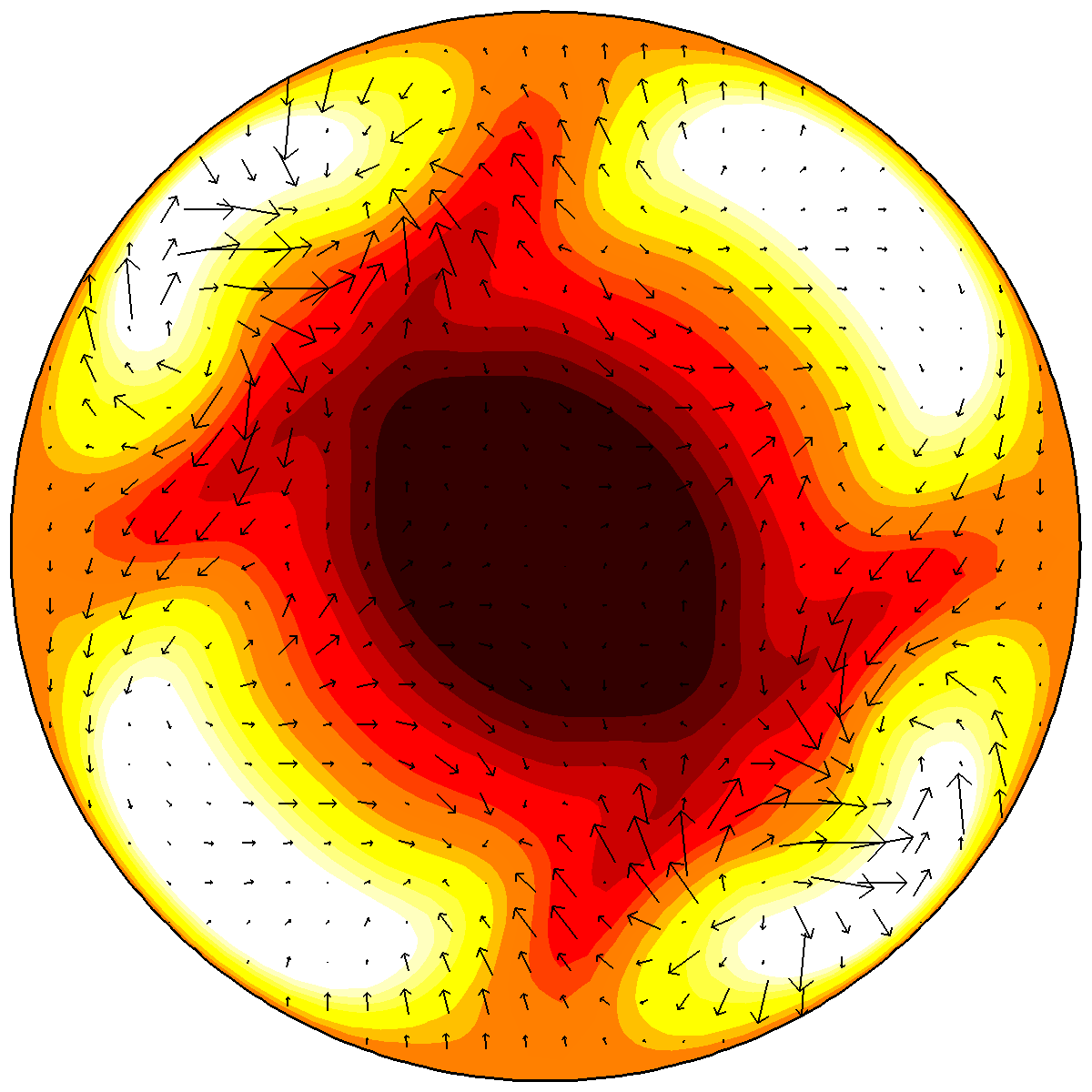}&
    \includegraphics[width=0.24\textwidth]{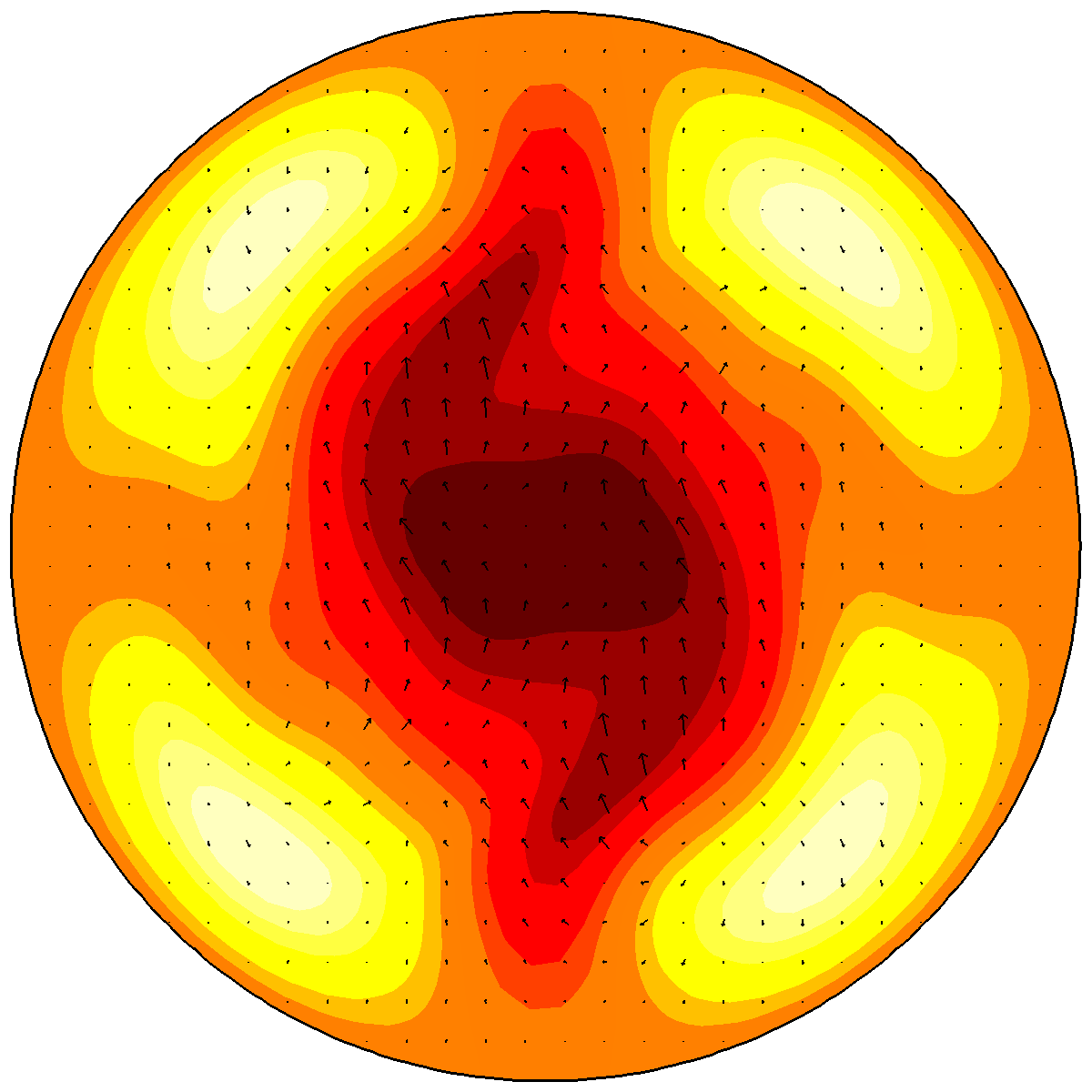}&
    \includegraphics[width=0.24\textwidth]{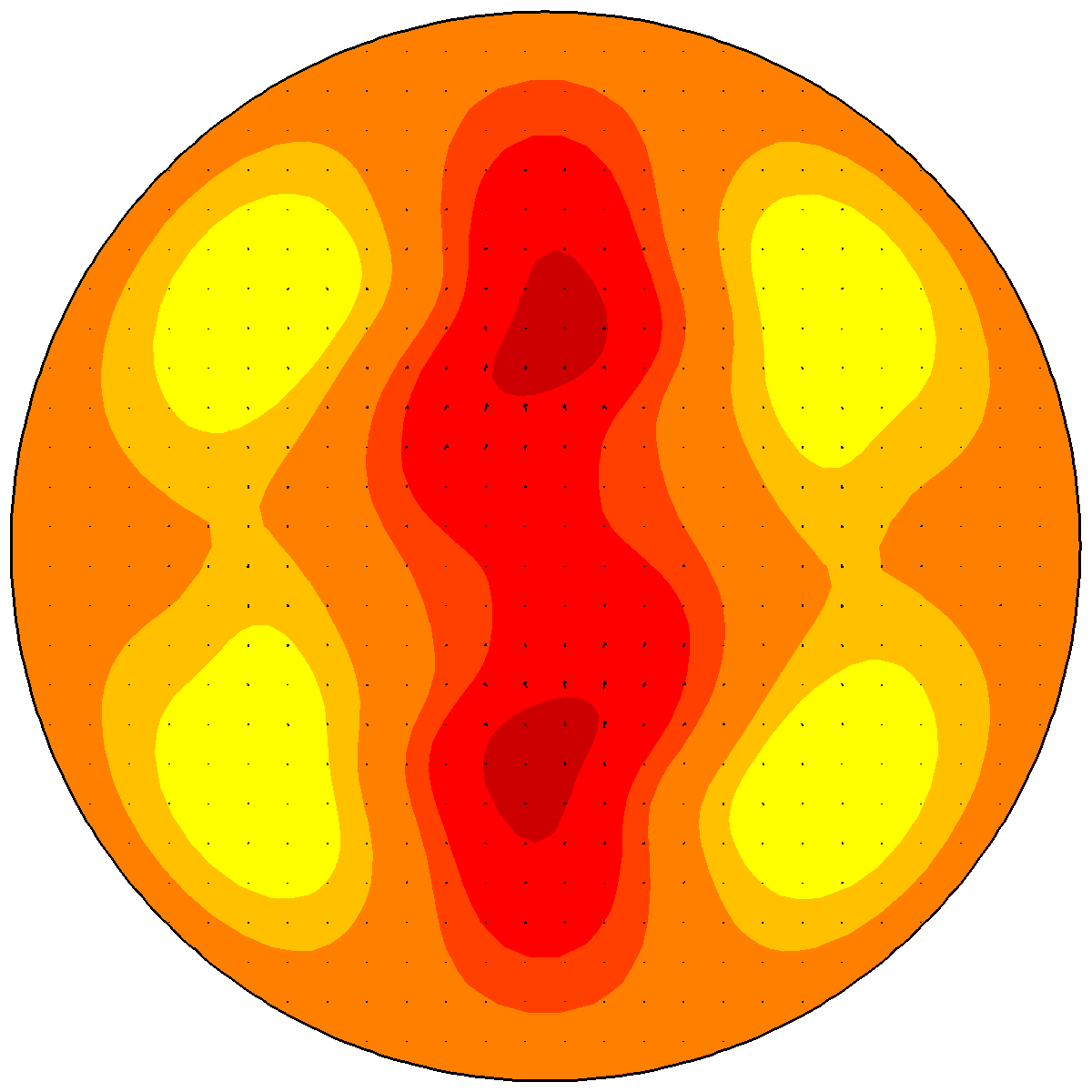}
  \end{tabular}
  \caption{ \label{fig:N2states}
    \Reqva\ for the cell \refeq{shortPipe}, reduced to \eqva\ by the
    \mslices. Colour map of stream-wise velocity relative to the laminar
    flow,
    lighter (darker)
    indicating positive (negative) $w$ in the range
    $[-0.6,0.7]$. The N2 states (LB, ML, MU and UB) have symmetries
    $(S,Z_2)$, where the symmetry $\Zn{2,\theta}$ is implied. 
    The S2 states originate from a symmetry-breaking
    bifurcation off the N2 branch and have symmetries $S$ and
    $\Zn{2,\theta}$ only. Shown is one fixed pipe section for each of the
    solutions. As the choice of the stream-wise position of such section
    is arbitrary, only meaningful comparison of different solutions is by
    their distance in the symmetry-\reducedsp.
    }
\end{figure}

For our first exploration of the state space of pipe flow, we have chosen
a cell size and Reynolds number combination \refeq{shortPipe} empirically
balanced so that $\Reynolds$ is just sufficient to sustain long periods
of turbulence. Among the many \reqva\ already known, we have chosen to
focus on the family of solutions classed N2 in \cite{Pringle09}.  At the
parameter values \refeq{shortPipe}, the N2 family has the upper and lower
branches (UB and LB), as well as two middle states (MU and ML), where
`upper' and `lower' refers to the friction or dissipation
\refeq{Power=I-D}, associated with each state. The middle states should
not be confused with the M branch of solutions documented in
\cite{Pringle09}, for which the M2 branch does not appear to exhibit
solutions at these particular parameters.  States of the S class do exist
at these parameter values, however. \refFig{fig:N2states} shows all
\reqva\ considered in this work.

\subsection{Sliced \reqva}
\label{s:eqbSols}

\begin{figure}
  \centering
(a)\includegraphics[width=0.45\textwidth]{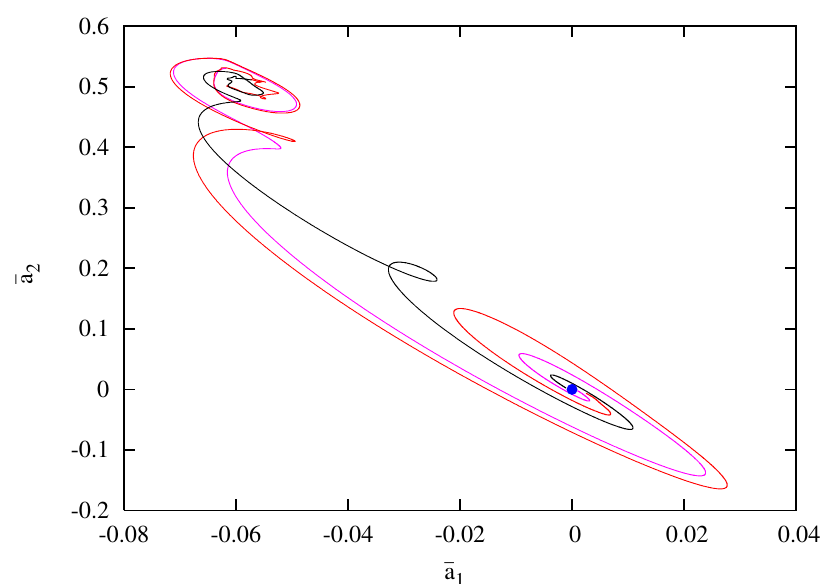}
(b)\includegraphics[width=0.45\textwidth]{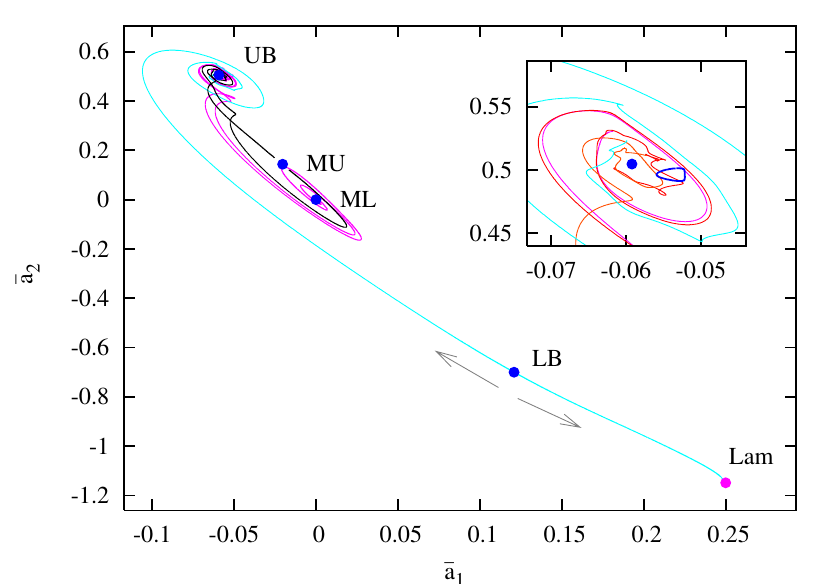}
  \caption{ \label{fig:M1loc2}
(a)
    Projection of the dynamics local to the ML {\reqv} which has been
    reduced to the \eqv\ at (0,0) within the \slice.  The local spiral of
    unstable trajectories is now clearly revealed, the ML state having
    only a single complex unstable eigenvalue within its $(S,Z_2)$
    symmetry subspace.
(b)
    All N2 equilibria, perturbations restricted to the $(S,Z_2)$
    symmetry subspace. The inset shows an expanded view near the UB
    state.  The dark blue loop is a tiny \rpo\ with period
    $\period{}=4.934$. Axes as in \refeq{eq:M1projdirs}.
  }
\end{figure}

For the sake of simplicity we consider first the dynamics restricted
to the $(S,Z_2)$ symmetry subspace of the N2 states. A convenient
property of the ML state at our parameter values \refeq{shortPipe} is
that it has only one complex unstable eigenvalue within this symmetry
subspace. The trajectories of small perturbations therefore spiral
away from ML as they follow its unstable manifold.  With ML drifting
in the axial direction, this local spiral would be difficult or
impossible to detect. Within the slice, however, the ML state is
reduced to an \eqv\ and the local spiral structure is clear, as shown
in \reffig{fig:M1loc2}\,({\it a}). To project onto the two dimensions
of the page, deviations from the ML state have been projected as in
\refeq{intrSspTraj}, against the real and imaginary components of its
complex stability eigenvector, $\hat{\be}_1$ and $\hat{\be}_2$
respectively,

\beq
\sspRed_i(\zeit) = \braket{\sspRed(\zeit)-\sspRed_\mathrm{ML}}{\hat{\be}_i}
\,.
\ee{eq:M1projdirs}
Once trajectories escape the neighbourhood of ML they are attracted to
another \statesp\ region
where the UB solution is to be found.
Applying the same projection to all N2 \reqva,
\reffig{fig:M1loc2}\,({\it b}) shows trajectories along their most unstable
directions. Shooting in opposite directions along the most unstable,
1-dimensional manifold of the MU state, one direction goes directly
towards the UB state, the other spirals around the ML state first.  All
trajectories within the $(S,Z_2)$ symmetry subspace are
attracted towards a region close to the UB state, where the dynamics
is mildly chaotic.  In this region we have
found a weakly unstable tiny \rpo\ of period $\period{}=4.934$, shown in the
inset to \reffig{fig:M1loc2}\,({\it b}), that appears to dominate the long-time
dynamics within the $(S,Z_2)$-invariant subspace. Shooting in
opposite directions from the LB state, trajectories proceed directly to
either the upper region or the laminar state, suggesting that LB \reqv\
lies within the laminar-turbulent boundary or `edge'.

\begin{figure}
  \centering
(a)\includegraphics[width=0.45\textwidth]{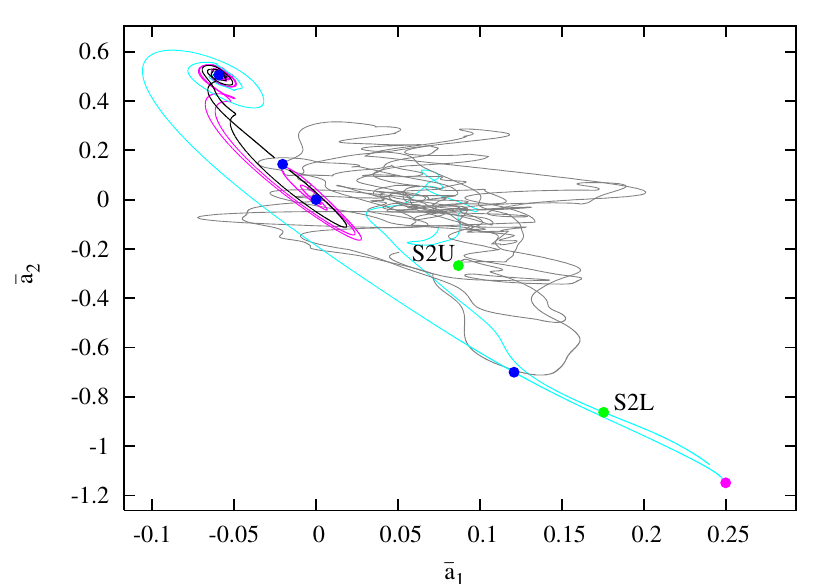}
(b)\includegraphics[width=0.45\textwidth]{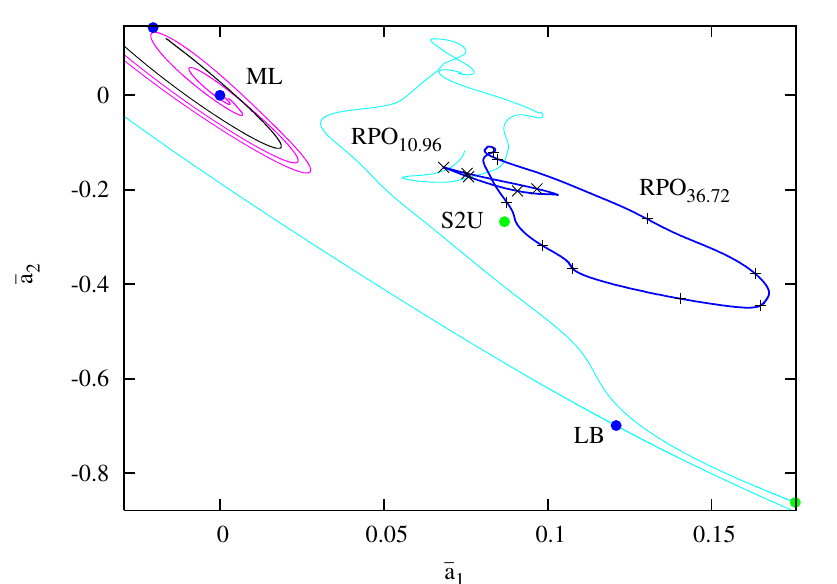}
  \caption{ \label{fig:M1FULL}
(a)
    Removal of the $Z_2$ symmetry of \reffig{fig:M1loc2} opens the
    system to far more chaotic, or `turbulent', transients (grey) within
    the $S$ symmetry solution space that appear to frequently visit the
    \reqv\ S2U, while the \reqv\ S2L appears embedded into the
    laminar-turbulent boundary.  The turbulent trajectory exhibits
    excursions to other states as well, most frequently ML and LB.
(b)
    Two \rpo s embedded within turbulence,  the same symmetry-\reducedsp\
    projection as \reffig{fig:M1loc2}. Crosses are spaced
    every $\Delta\zeit = 2 D/U$ on \rpo\ $\RPO{10.96}$ and pluses are spaced 4$\,D/U$ on
    \rpo\ $\RPO{36.72}$. On this scale the tiny $\RPO{4.934}$
    [\reffig{fig:M1loc2}\,({\it b}), inset] only explores a region about
    the size of the plot dots, and plays no role in turbulent dynamics.
  }
\end{figure}

Without restriction of dynamics to the $(S,Z_2)$-invariant
subspace, trajectories stray much further from ML and show turbulent
behavior. In order to track such trajectories all {\reqva} states
listed in \reftab{tab:RPOs} were deployed as templates,
$\slicep{}^{(j)}$, $j=1,2,\cdots,6$, whereas the single
\template\ point $\slicep = \ssp_{\mathrm{ML}}(0)$ sufficed for the
symmetry reductions within the $(S,Z_2)$ subspace. Switching from
one local slice to the next nearest one keeps the {\phaseVel}
\refeq{reconstrEq} finite (see \reffig{fig:sliceimage}{\it b}) and
enables tracking of turbulent trajectories in the \reducedsp.

A typical trajectory is shown in \reffig{fig:M1FULL}\,({\it a}).  Within
the ($S$,$Z_2$)-invariant subspace trajectories hover near the UB
state; but when $Z_2$ symmetry is relaxed, and only $S$ symmetry is
enforced, the trajectories explore a far greater region of \statesp, and
appear to be representative of turbulence in the full \statesp.  The
neighbourhood of the S2U state is visited frequently, and excursions to
other states are occasionally seen. Interestingly, an excursion is
observed towards the LB state. Its attracting manifold therefore appears
to penetrate into the turbulent region, and, as it lies on the laminar
turbulent boundary, attraction towards this manifold may be responsible
for the observed sudden relaminarisation events.

\subsection{Relative periodic orbits in pipe flow}
\label{s:rpos}

\begin{table}
   \centering
   \begin{tabular}{lcllllllllllll}  
       &&$\frac{\timeAver{E}}{E_\mathrm{lam}}$
       & $\frac{\timeAver{D}}{D_\mathrm{lam}}$  
       & $~~\timeAver{\velRel}$
				   &  ~~$\shift$
                                   & ~~$\period{}$
                                   & \# unst.
                                   & $\eigRe[max]  \pm i\,\eigIm[max]$\\
   \hline
   symmetry & $(S,Z_2)$ \\
   LB && 0.94330 & 1.2137 & 1.551 &&&   1r &  0.07906 \\
       & $S$    &&       &       &&&  +0    &  0.07906\\
   ML && 0.88662 & 1.6974 & 1.421 &&&   1c &  0.02490 $\pm i$\,0.07323\\
       & $S$    &&       &       &&&  +1r+2c & 0.2704 $\pm i$\,1.515 \\
   MU && 0.87723  & 1.8322 & 1.394 &&&   1r &  0.05617  \\
       & $S$    &&       &       &&&  +1r+2c        &  0.3267 $\pm i$\,1.543\\
   UB && 0.85273  & 2.4990 & 1.298 &&&   3c &  0.2179 $\pm i$\,1.983  \\
       & $S$    &&       &       &&&  +6c           &  0.4231 $\pm i$\,1.660\\
   $\RPO{4.934}$&& 0.85137 & 2.4451 & 1.302 & ~6.423  & ~4.934
                                   &   1c &  0.1242 $\pm i$\,0.3819  \\
       & $S$    &&       &       &&&  +6c           &  0.4417 $\pm i$\,0.3284  \\
   \hline
   symmetry &  $S$ \\
   S2U&& 0.89383 & 1.4495   & 1.296 &&&  1c         &  0.05592 $\pm i$\,0.5215 \\
   S2L&& 0.96159 & 1.1191  & 1.522 &&&  1r         &  0.1090  \\
   $\RPO{10.96}$&& 0.88845 & 1.5205 & 1.265 & 13.868 & 10.96
                                    &  1r+2c &     0.06051 $\pm i$\,0.15383 \\
   $\RPO{36.72}$&& 0.89515 & 1.4865 & 1.291 & 47.417 & 36.72
                                    &  2r+5c & 0.08636 $\pm i$\,0.0900 \\
   ergodic      && 0.8787 & 1.671 & 1.274 &   &
                                    &  & $\approx 0.11$
   \end{tabular}
   \caption{\label{tab:RPOs}
      All \reqva\ and \rpo s studied in this paper for pipe \refeq{shortPipe},
      split by solution symmetry:
      (mean) kinetic energy $\timeAver{E}$;
      (mean) dissipation $\timeAver{D}$, both in laminar solution units;
      \reqv\ downstream {\phaseVel} $\velRel$ or \rpo\ mean {\phaseVel}
      $\timeAver{\velRel}_p= \shift_p/\period{p}$;
      accumulated \rpo\ shift $\shift_p$ (not modulo the periodic cell length $L=5.0265\ldots$);
      period $\period{p}$; 
      the number of unstable eigen-directions within the solution's
      symmetry subspace, r$\,=\,$real, c$\,=\,$complex;
      the leading Floquet exponent $\eigExp[j]= \eigRe[j] \pm i\eigIm[j]$.
      For the upper part of the table,
      numbers in $S$ rows are for symmetry-breaking eigenvalues
      when $Z_2$ is removed.
      `Ergodic' refers to long time average computed from evolution
      of typical turbulent states.
   }
\end{table}

\begin{figure}
   \centering
   \includegraphics[width=0.7\textwidth]{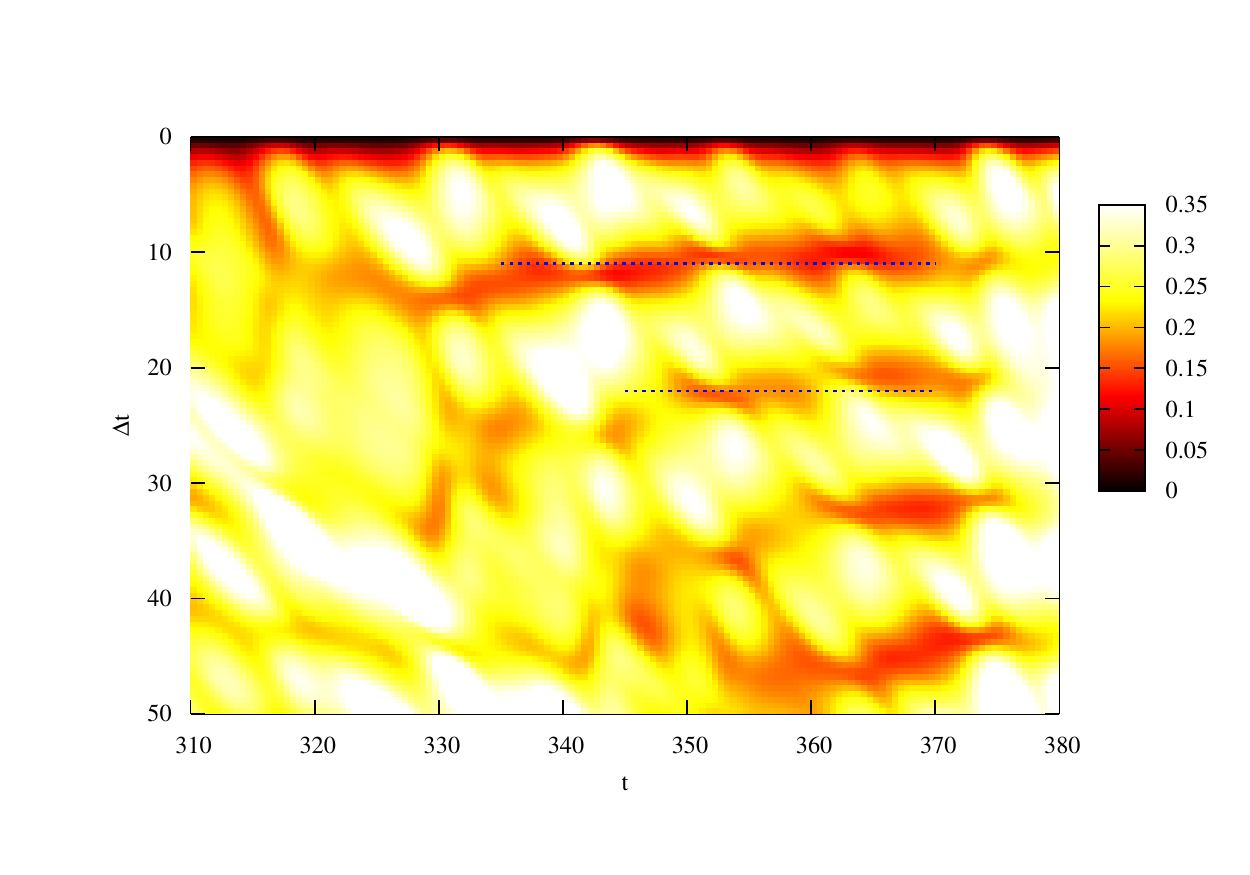}
   \caption{ \label{fig:NormDiff}
      Search for recurrences within the \slice.  Each state is compared
      with the state at earlier times $\Delta t$ before, shading
      indicates the relative distance
      $\Norm{\sspRed(\zeit)-\sspRed(\zeit-\Delta t)}_c/\Norm{\sspRed(\zeit-\Delta t)}_c$.
      The minima
      indicated by the horizonal lines suggest that an orbit of period
      $T\approx 11$ is shadowed for $\zeit \leq 40$
        (all times are expressed in units of $D/U$).
      Newton search
      indeed confirms this, by finding there the weakly unstable $\RPO{10.96}$. Note that
      the tiny \rpo\ $\RPO{4.934}$ from \reffig{fig:M1Orb}\,({\it a}) lies far
      from the turbulent region, and thus does not show up in recurrence
      plots.
   }
\end{figure}

Without symmetry reduction, the detection of a recurrence, i.e.\ that
current state is close in structure to an earlier state on the same
trajectory, requires calculating the minimum distance between their
group orbits, \ie, minimum over all possible shifts.
Within the symmetry-\reducedsp\ the determination of
recurrences is simple --- by construction,
a slice is the set of all nearby group orbit states closest to a given
template, with symmetry shifts quotiented out, hence
all group orbits are reduced to points, and all \rpo s to \po s.
The shifts $\shift_p$ are determined by the slice condition \refeq{eq:slcond}.
\refFig{fig:NormDiff} shows a recurrence plot used to detect the signal
of a turbulent trajectory that shadows a nearby \rpo.  The indicated
minimum at $\Delta t\approx 11$ and its repeats are seen for a while as
the \rpo\ $\RPO{10.96}$ is shadowed for a rather long time. States from
this minimum, along with the relative stream-wise shift for the candidate
trajectory, $\shift(\zeit)-\shift(\zeit-\Delta t)$, were passed to our
Newton--Krylov code.  This led to the discovery of the \rpo\ that we
label $\RPO{10.96}$, and another recurrence plot led to the \rpo\
$\RPO{36.72}$, both plotted in \reffig{fig:M1FULL}\,({\it b}).
(In absence
of a systematic symbolic dynamics, we label $\RPO{\period{}}$ by its
period $\period{}$.)

\begin{figure}
   \centering
(a)\includegraphics[width=0.45\textwidth]{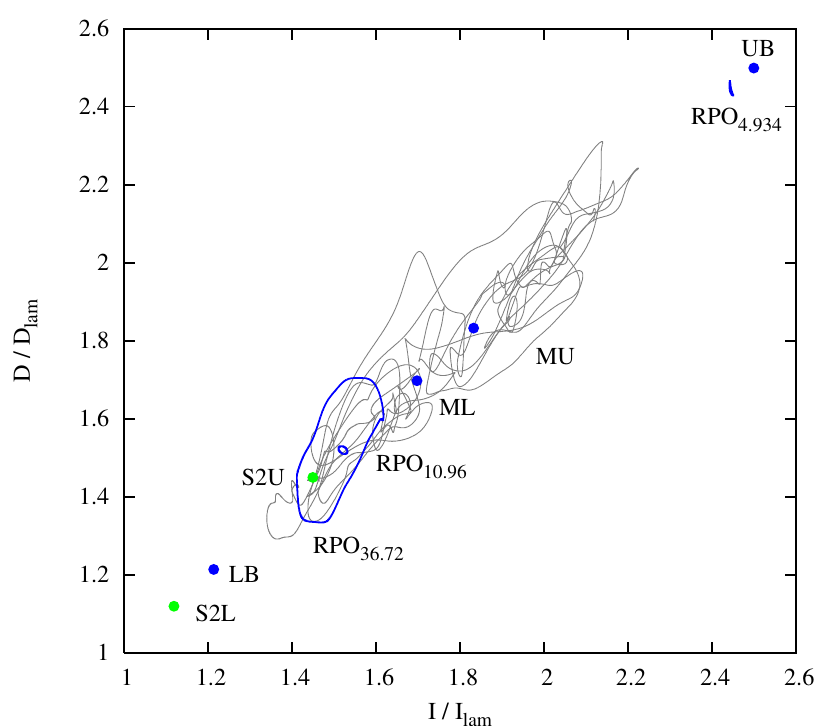}
(b)\includegraphics[width=0.45\textwidth]{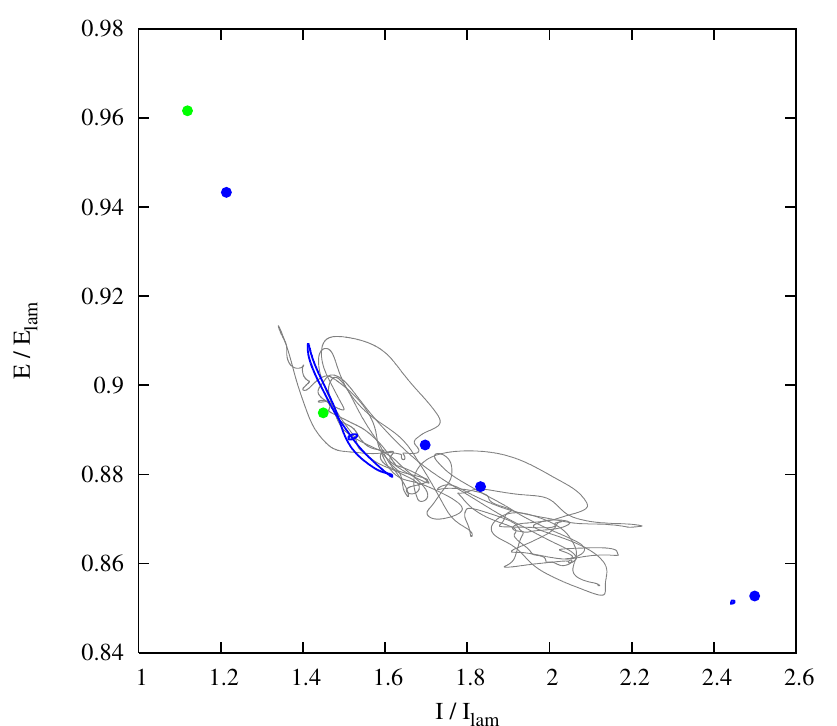}
   \caption{ \label{fig:M1Orb}
Rate of the energy input at the walls $I$ versus
   (a) the dissipation rate $D$, and
   (b) the energy $E$, see \refeq{Power=I-D}.
   Plotted are all invariant solutions of
    \reftab{tab:RPOs}, together with a typical turbulent orbit. The
    tiny $\RPO{4.934}$ is visible as a little twiddle just below UB
    \reqv. On the scale of these plots, the \cite{duguet08} \rpo\ would
    be indistinguishable from its mother \reqv.
   }
\end{figure}

\begin{figure}
\centering
\begin{tabular}{cccc}
  $t=0$ & $T/4$ & $T/2$ & $3T/4$\\
  \includegraphics[width=0.24\textwidth]{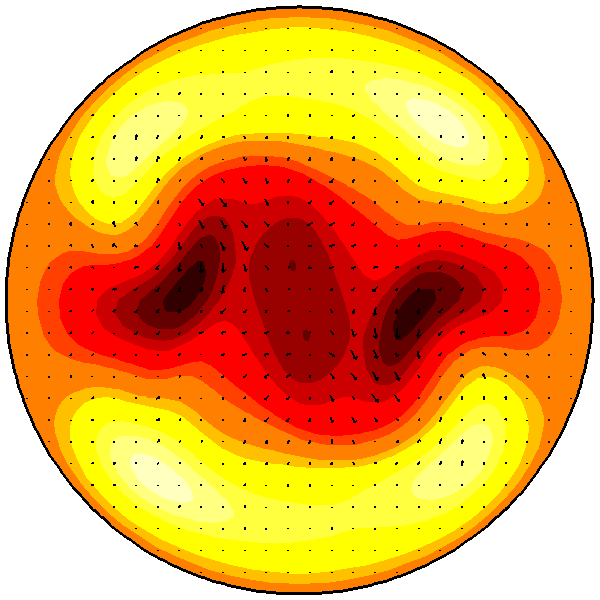}&
  \includegraphics[width=0.24\textwidth]{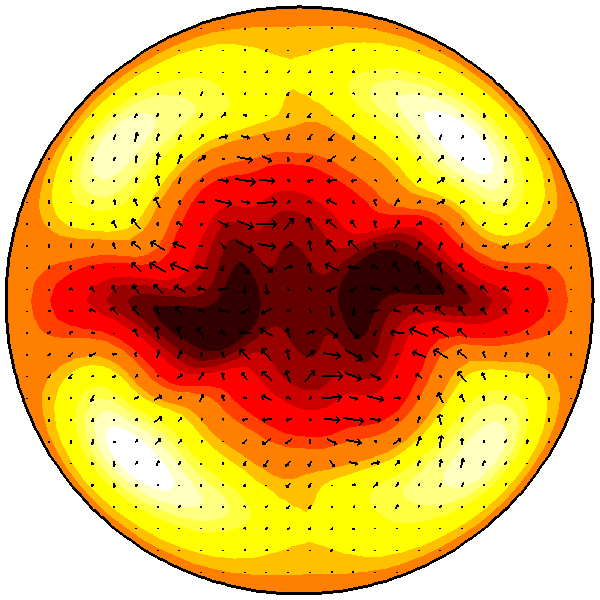}&
  \includegraphics[width=0.24\textwidth]{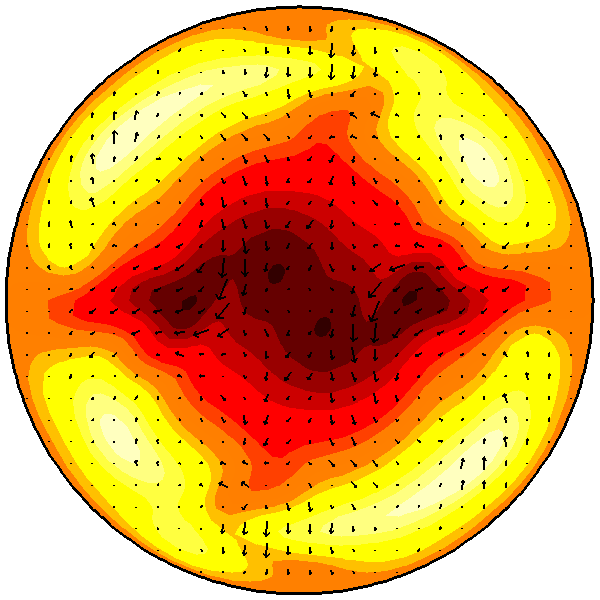}&
  \includegraphics[width=0.24\textwidth]{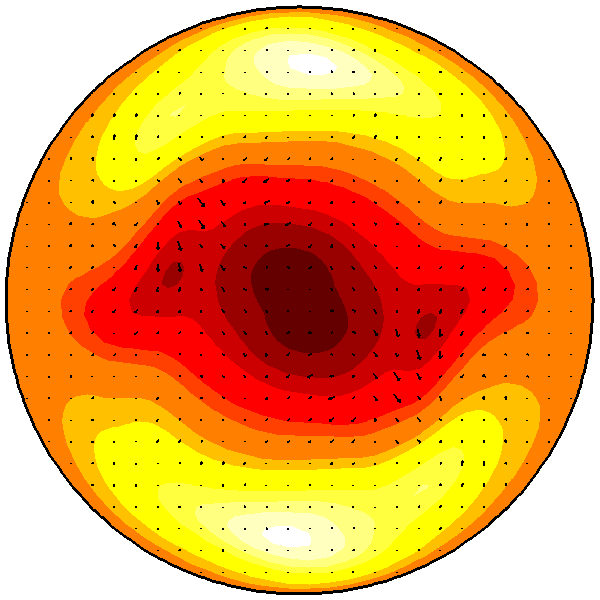}
\end{tabular}
\caption[First bursting \rpo]{
  Four snapshots of \rpo\ $\RPO{36.72}$ at the same fixed pipe section as
  in \refeq{fig:N2states}, reduced by the \mslices\ into a \po\ solution
  (see movie \HREF{../movie/rpo\_burst1.flv}{online}). A colormap of
  stream-wise velocity is shown, with white (black) indicating positive
  $w=0.6 \, U$ (negative $w=-0.7 \, U$) velocity with respect to laminar
  flow. The $\zeit=0$ state has been chosen to be the closest passage (in the
  energy norm) to \reqv\ S2U, see \refeq{fig:N2states}.
}
\label{f:rpo_burst1}
\end{figure}

Several two-dimensional projections of \rpo s $\RPO{36.72}$,
$\RPO{10.96}$ and the tiny $\RPO{4.934}$ are given in
\reffig{fig:M1Orb}, along with the N2 and S2 states used in this paper
(same colour coding as previous plots). Figure~\ref{f:rpo_burst1} (see
movie online) shows flow snapshots of $\RPO{36.72}$ at a fixed axial
cross-section. The movie has been taken after reducing the continuous
symmetry with the method of the slices and hence shows how the orbit
closes after one period. The orbit consists of a slow nearly quiescent
phase, during which the neighbourhood of S2U is visited, followed by a
period of intense turbulent bursting. This behaviour suggests that
the orbit $\RPO{36.72}$ may be related to a global homoclinic bifurcation
off S2U.

\subsection{Discussion}
\label{s:discuss}

Symmetry reduction by the \mslices\ in a high-dimensional flow thus
reveals dynamics around recently discovered \reqva, both local and
global, and leads to the discovery of first \rpo s in pipe flow that, as
they have been extracted from turbulent trajectories, can be expected to
be dynamically important. While the tiny \rpo\ $\RPO{4.934}$ appears to
originate from a Hopf bifurcation off a \reqv\ very nearby \citep[have
also found similar local \rpo s]{duguet08,mellibovsky11} $\RPO{10.96}$
and $\RPO{36.72}$ exhibit temporal variation typical of the turbulence
found in our computational domain \refeq{shortPipe}.

Visualisations of physical quantities, such as in \reffig{fig:M1Orb},
are often used in the literature to infer the importance of coherent
solutions (\eg \reqva\ and \rpo s in pipe flow) in turbulent
flow. Here the dissipation $D$ and input $I$ of \reffig{fig:M1Orb}(a),
for example, clearly show that states S2L, LB and UB are far from the
turbulent flow. However, the converse is usually not true. As the
energy balance \refeq{Power=I-D} forces all \reqva\ and the averages
over all \rpo s, and turbulent flow, to lie on the diagonal $I=D$,
\reqva\ that feature frictions close to the turbulent average may
appear to be in core of the turbulent region. For example,
\reffig{fig:M1Orb} suggests that the ML and MU states may be
representative of the turbulent dynamics. The projection within the
slice (see \reffig{fig:M1FULL}) reveals that in fact these two states,
despite having the `right friction', are far from the turbulent
dynamics in phase space. Our results show that the neighbourhoods of
known \reqva\ (\reftab{tab:RPOs}), with the exception of S2U, are
visited only for a small fraction of time, in agreement with earlier
estimates \citep{KeTu06,SchEckVoll07,WillKer08}. It is only the
\reducedsp\ projections that make it quite clear that only
$\RPO{10.96}$, $\RPO{36.72}$ and S2U are embedded in the region
associated with turbulence. To sum up, phase-portraits using
physically motivated quantities such as dissipation, input and kinetic
energy, may be used to rule out the relevance of coherent states in
turbulent flow but not to confirm their relevance.

Another important point is that determination of \reqva\ by
bifurcations and continuation is often physically misleading. Almost
all of the previously found \reqva\ and nearby tiny \rpo s are highly
unstable to perturbations out of their symmetry subspace, highly
repelling and not participants in the asymptotic dynamics (see
\reftab{tab:RPOs}). The exception to this are the nearly laminar
lower-branch states, which play a key role organising the dynamics of
the turbulent-laminar boundary or edge. Instead, recurrences in
turbulent flow used as initial guesses for Newton--Krylov methods
allow it to find solutions that are relevant to the turbulent
dynamics. Note that the new \rpo s revealed here are associated to the
lower dissipation region of turbulent flow; but is expected that
application of the method slices together with a systematic study of
recurrences in the upper region will yield new `turbulent' \rpo s and
\reqva.

Finally, it is worth emphasising that restriction of dynamics to
flow-invariant subspaces can potentially be very misleading. For
example, \reffig{fig:M1loc2} and \ref{fig:M1FULL} exhibit completely
different dynamics. In this case, imposing the rotate-and-reflect
symmetry, in addition to shift-and-reflect, results even in the
absence of turbulent dynamics.  Hence, despite the similarity of the
dynamics of full-space and shift-and-reflect turbulence, our choice
may also be problematic.

\section{Conclusion and perspectives}
\label{s:concl}

As a turbulent flow evolves, every so often we catch a glimpse of a
familiar structure. For any finite spatial resolution, the flow stays for
a finite time in the neighbourhood of a coherent structure belonging to an
alphabet of admissible fluid states, represented here by a set of \reqv\
and \rpo\ solutions of \NS. These are not the `modes' of the fluid; {they
do not provide a decomposition of the flow into a sum of components at
different wavelengths, or a basis for low-dimensional modelling.} Each
such solution spans the whole range of physical scales of the turbulent
fluid, from the outer wall-to-wall scale, down to the viscous dissipation
scale. Numerical computations require sufficient resolution to cover all
of these scales, so no {global} dimension reduction is likely. The role
of invariant solutions of \NS\ is, instead, to partition the
$\infty$-dimensional \statesp\ into a finite set of neighbourhoods visited
by a typical long-time turbulent fluid state.

Motivated by the recent observations of \recurrStr s in experimental and
numerical turbulent flows, we initiated here an exploration of the
hierarchy of \reqva\ and \rpo s of fully-resolved transitionally
turbulent pipe flow in order to describe its spatio-temporally chaotic
dynamics.
For pipe flow \reqva\ and \rpo s embody a vision of turbulence as a
repertoire of recurrent spatio-temporal coherent structures explored by
turbulent dynamics. The new \rpo s that we present here are a part of the
backbone of this repertoire. Given a set of invariant solutions, the next
step is to understand how the dynamics interconnects the neighbourhoods of
the invariant solutions discovered so far.
Currently, a taxonomy of these myriad states eludes us, but emboldened by
successes in applying periodic orbit theory to the simpler \KS\ problem
\citep{Christiansen97,lanCvit07,SCD07}, we are optimistic.

The reader might rightfully wonder what the short pipe periodic cells
studied here and in \pCf\ have to do with physical, wall-bounded shear
flows in general, with large aspect ratios and physical boundary
conditions?
The $3D$ fluid states captured by the short pipe invariant solutions and
their unstable manifolds are strikingly similar to states observed both
in experiments and in numerical simulations of longer pipes
\citep{science04}, while the turbulent dynamics visualised in \statesp\
appears to be pieced together from close visitations to \cohStr s
connected by
transient interludes. Nevertheless, one of the outstanding issues that
must be addressed in future work is the small-aspect cell periodicities
imposed for computational efficiency. In case of the pipe flow, most
computations of invariant solutions have focused on stream-wise periodic
cells barely long enough to allow for sustained turbulence. Such small
cells introduce dynamical artifacts such as lack of structural stability
and stream-wise cell-size dependence of the sustained turbulence states.
Here we can draw inspiration from pattern-formation theory, where the
most unstable wavelengths from a continuum of unstable solutions set the
scales observed in simulations, with recent progress reported both from
the `microscopic scales' \citep{SGB10}, as well as long pipe experiments
and phenomenology \citep{AMdABH11}.

The main message of this paper is that if a problem has a continuous
symmetry, the symmetry \emph{must} be used to simplify it. Ignore it at
your own peril, as has been done earlier in \KS\ \citep{Christiansen97}
and \pCf\ \citep{GHCW07}; the invariant solutions found by restricting
searches to the discrete-symmetry invariant subspaces have little if
anything to do with the full \statesp\ explored by turbulence, no more
than the \eqv\ points of the Lorenz flow have to do with its strange
attractor.
Note also that the shift of a pipe flow into a \slice\ {\em is not}
a stream-wise average over the 3D pipe flow.  It is the full
flow snapshot, embedded in the $\infty$-dimensional \statesp. Symmetry
reduction is not a dimensional-reduction scheme, or flow modelling by
fewer degrees of freedom: the \reducedsp\ is also $\infty$-dimensional
and no information is lost, one can go freely between solutions in the
full and reduced \statesp s by integrating the associated
{\em reconstruction equations}.

Symmetry reduction by \mslices\ is numerically efficient.
Coupled with our \statesp\ visualisations, it allows for explorations of
high-dimensional flows that were hitherto unthinkable.
Symmetry reduction is here achieved, and now all pipe flow solutions can
be plotted together, as one happy family: all points equivalent by
symmetries are represented by a single point, families of solutions are
mapped to a single solution, \reqva\ become \eqva, and \rpo s become \po
s.
Without symmetry reduction, no full understanding of pipe and plane
\pCf s is possible.

\begin{acknowledgments}
We would like to acknowledge R.~R.~Kerswell for providing \reqva\
solutions data.
We are indebted to
R.~L.~Davidchack,
S.~Froehlich,
B.~Hof,
and
E.~Siminos
for inspiring discussions,
and
D.~W.~Spieker for contributing to the symmetry classification
of \refsect{s:symm}.
A.~P.~W.\ was initially funded by the European
Community's Seventh Framework Programme FP7
2007-2013 under Grant agreement No. PIEF-GA-2008-219-233.
M.~A.\ was supported by the Max-\-Planck-\-Gesellschaft.
P.~C.\ thanks G.~Robinson,~Jr.\ for support, and
 Max-Planck-Institut f\"ur Dynamik und Selbstorganisation,
G\"ottingen for hospitality.
P.~C.\ was partly supported by NSF grant DMS-0807574
and
2009 Forschungspreis der Alexander von Humboldt-Stiftung.
\end{acknowledgments}

\appendix
\section{Discrete symmetries}
\label{appe:DiscSymmPipe}

In addition to azimuthal reflection, invariant solutions can exhibit
further discrete symmetries that derive from azimuthal and stream-wise
periodicities over the computational cell \refeq{cellPipe}.

Periodicity in the azimuthal direction allows for solutions with discrete
cyclic symmetry $\LieEl(2\pi/m, 0)$, defined for integer $m$.
Velocity fields invariant under such rational azimuthal shifts
are said to be invariant under the discrete cyclic group $\Zn{m,\theta}$.
Note that all solutions are invariant under $\Zn{1,\theta}$, and
given the assumed stream-wise periodicity, under $\Zn{1,z}$ as well.
This permits the study of states in the reduced computational cells
$\bCell=[0,1/2]\times[0,2\pi/m]\times[0,\pi/\alpha]$, where $L=\pi/\alpha$.
Calculations in larger domains are required to determine subharmonic bifurcations.

Consider states invariant under $\Zn{m,\theta}$ and $\Zn{1,z}$, and
denote half-shifts within our reduced cell, in $\theta$ and $z$ respectively,
by $\LieEl_\theta=\LieEl(\pi/m,0)$ and $\LieEl_z=\LieEl(0,L/2)$.
For the special case of a
half-shift in azimuth, $\sigma$ and $\LieEl_\theta$ commute so that
\beq
G = \Dn{1} \times \Zn{m,\theta} \times \Zn{1,z} \subset \Gpipe
\ee{Gpipe8sbgrp}
is abelian and of order 8,
\beq
   G = \{e,\LieEl_\theta,\LieEl_z,\LieEl_\theta\LieEl_z,
    \sigma,\sigma\LieEl_\theta,\sigma\LieEl_z,\sigma\LieEl_\theta\LieEl_z\} .
\eeq
Focus lies on the following subgroups:
\beq
   Z = \{e,\sigma\}, \qquad
   S = \{e,\sigma\LieEl_z\}, \qquad
   \Omega_m \ = \{e,\LieEl_\theta\LieEl_z\}
 \,.
\ee{ShiftRefl}
The first is the `reflectional', or `mirror' symmetry,
the second is the `shift-and-reflect' symmetry, and
the third is the `shift-and-rotate' symmetry.
States invariant under $\LieEl_\theta$ or $\LieEl_z$ are
invariant under $\Zn{2m,\theta}$ or $\Zn{2,z}$ and hence become redundant
upon redefinition $m := 2m$ or $\alpha := 2\alpha$ (\ie\ they reduce to 
half-cells).
It can also be shown that
$\sigma\LieEl_\theta=\LieEl_\theta^{-1/2}\sigma\LieEl_\theta^{1/2}$,
where $\LieEl_\theta^{1/2}$ is the half-half-shift,
and therefore that $\sigma\LieEl_\theta\LieEl_z=\LieEl_\theta^{-1/2}
\sigma\LieEl_z\,\LieEl_\theta^{1/2}$.  Invariance under these combinations
is conjugate to $Z$ and $S$.  We use, however, the `rotate-and-reflect'
subgroup, denoted by
\beq
Z_m=\{e,\sigma\LieEl_\theta\}.
\eeq
which has mirror reflection planes located at 
$\theta=\pm\pi/(2m)$ (see \reffig{fig:N2states} for the case $m=2$).

The first \reqva\  found for pipe flow were invariant under 
$S$ and $\Zn{m,\theta}$ for $m=2,3,4,...$ \citep{FE03,WK04}.  
More recently 
the `missing' $m=1$ state has been located \citep{Pringle07} and many more
states invariant under more than one of the above classes
\citep{Pringle09}. States invariant under $(S,Z)$
implies invariance under $\sigma\sigma\LieEl_z=\LieEl_z$, and
hence under $\Zn{2,z}$, reducing to the half-length pipe. Invariance
under $(S,\Omega_m)$ is permissible, however, and
using the combinations above it can be calculated that 
$(S,\Omega_m)=(S,Z_m)=(Z_m,\Omega_m)$.
Such states have been termed `highly symmetric' by 
\cite{Pringle09}.
As reflection is arguably easier to visualise than shift-and-rotate,
we use the notation $(S,Z_m)$ for these states.


\bibliographystyle{jfm}
\bibliography{../bibtex/pipes}

\end{document}